\documentclass[12pt]{article}
 \usepackage[nottoc]{tocbibind}
\usepackage[colorlinks=true,linkcolor=blue,urlcolor=magenta, citecolor=blue]{hyperref}
\usepackage{fullpage}
\usepackage{amssymb,graphicx}
\usepackage{amsmath}
\usepackage[margin=0.5cm]{caption}
\usepackage{hyperref} 
\usepackage{cite}
\usepackage{upgreek}

\def\e{{\rm e}}

\def\eps{\epsilon}
\def\d{\partial}
\def\l{\left(}
\def\r{\right)}

\newcommand{\be}{\begin{equation}}
\newcommand{\ee}{\end{equation}}
\newcommand{\bea}{\begin{eqnarray}}
\newcommand{\eea}{\end{eqnarray}}
\newcommand{\bg}{\begin{gather}}
\newcommand{\eg}{\end{gather}}
\newcommand{\bseq}{\begin{subequations}}
\newcommand{\eseq}{\end{subequations}}

\def\half{\frac{1}{2}}

\begin{document}
\baselineskip=15.5pt
\begin{titlepage}
\begin{center}
{\Large\bf  Towards a Theory of the QCD String}\\
\vspace{0.5cm}
{ \large
Sergei Dubovsky$^{\rm a}$ and Victor Gorbenko$^{\rm b}$
}\\
\vspace{.45cm}
{\small  \textit{   $^{\rm a}$Center for Cosmology and Particle Physics,\\ Department of Physics,
      New York University\\
      New York, NY, 10003, USA}}\\ 
      \vspace{.1cm}
{\small  \textit{  $^{\rm b}$Stanford Institute for Theoretical Physics,\\Department of Physics, Stanford University, Stanford, CA 94305, USA }}\\ 
\end{center}
\begin{center}
\begin{abstract}
We construct a new model of four-dimensional relativistic strings with integrable dynamics on the worldsheet.
In addition to translational modes  this model contains a single {\it massless} pseudoscalar 
worldsheet field---the worldsheet axion.  The axion couples to a topological density which counts the self-intersection number of a string. The corresponding coupling is fixed by integrability  to $Q=\sqrt{7\over 16\pi}\approx 0.37$.
 We argue that this model is a member of a
larger family  of  relativistic non-critical integrable string models. 
 This family includes and extends conventional non-critical strings described by the linear dilaton CFT.  Intriguingly, recent lattice data in $SU(3)$ and $SU(5)$ gluodynamics reveals the presence of a {\it massive} pseudoscalar  axion on the worldsheet of confining flux tubes. The value of the corresponding coupling, as determined from the lattice data, is equal to $Q_L\approx0.38\pm0.04$.
\end{abstract}
\end{center}
\end{titlepage}
\tableofcontents
\newpage
\section{Introduction}
Integrable relativistic quantum field theories are among the most beautiful creatures in a zoo of mathematical physics, with spectacular applications to real world systems, such as cobalt niobate (CoNb$_2$O$_6$)
\cite{coldea2010quantum}.
However, up until recently, integrability was confined to the realm of $(1+1)$-dimensional physics. For various reasons (most notably, due to the Coleman--Mandula theorem), one might have been skeptical about existence of integrable higher-dimensional  quantum field theories. This all changed with the discovery of integrability in the planar ${\cal N}=4$ supersymmetric Yang--Mills (SYM), as reviewed in \cite{Beisert:2010jr}. Even though it is presently unclear what is the precise definition of integrability in a higher-dimensional field theory, there is an overwhelming evidence that the planar  ${\cal N}=4$ SYM does satisfy the most important criterion---it is a non-trivial solvable model. 

Given this success it is natural to ask what are other examples, if any, of integrable higher-dimensional models. In particular, one may wonder whether integrability is possible in a non-conformal (gapped) 
higher-dimensional theory. The first step in addressing this question should be to make it more precise what one understands by integrability.  Let us propose a possible way to achieve this, applicable in a broad class of theories.

Namely,
consider a gapped confining gauge theory with a center symmetry or some analogous property, which ensures that confining strings (flux tubes) cannot break. Pure gluodynamics is the most familiar prototype example of such a theory. 
Then we can consider a sector of this theory with a single infinitely long flux tube stretching through all of the space. Alternatively, one may compactify the theory on a large cylinder and consider a sector with a flux tube wound around the compact dimension. The RG flow describing the theory in this sector interpolates between a four-dimensional description at high energies, and a two-dimensional worldsheet theory describing the flux tube dynamics at low energies. The  low energy two-dimensional theory necessarily contains ``branons" $X^i$'s --- gapless Goldstone modes associated  with spontaneous breaking of the bulk Poincare group due to the presence of a string.
In addition, it may contain other light (or even massless) modes.

Given we understand better what integrability means in a two-dimensional context, a natural question to ask is whether it is possible to construct examples where the worldsheet theory is integrable.
To be  more precise, note that at energies below the mass of the lightest bulk particle (the lightest glueball in the pure glue case), $E<m_g$,
the worldsheet dynamics is described by a unitary two-dimensional $S$-matrix.
Then the question is whether this $S$-matrix may be factorizable, so that there is no particle production on the worldsheet. 

This criterion for integrability also matches well with the ${\cal N}=4$ example. Being a conformal theory, ${\cal N}=4$ SYM does not exhibit confining flux tubes. However, strings propagating in the dual $AdS_5\times S_5$ geometry provide their close analogue.  Currently, the best understanding of ${\cal N}=4$ SYM integrability is through the integrability of the Green--Schwartz description of these dual strings, which is analogous to what we are looking for for confining flux tubes.

Most likely, it is too much to ask for the worldsheet theory to be integrable at finite number of colors $N_c$, and just like in the ${\cal N}=4$ case one should look for integrability in the planar limit. In fact, at large $N_c$ the regime of validity of the two-dimensional description extends to energies even higher than $m_g$ (see Fig.~{\ref{fig_scales}). Indeed, at large $N_c$ the probability to emit a bulk glueball in a collision of  long string excitations is suppressed by a factor of $1/N_c^2$, so that the worldsheet theory is decoupled from the bulk up to a parametrically higher energy scale $E\sim \Lambda_{N_c}\gg m_g\sim\ell_s^{-1}$, where $\ell_s$ is the string length. 
As discussed in more details in the concluding Section~\ref{sec:last},  the decoupling scale $\Lambda_{N_c}$ is likely to exhibit a power-law growth in the planar limit, $\Lambda_{N_c}\ell_s\to \infty$.
Then, given that the spectrum of low-lying glueballs has a good large $N_c$ limit, one expects that the planar limit produces an $N_c$-independent  UV complete two-dimensional theory on the worldsheet\footnote{Barring possible subtleties with exchanging high energy and infinite $N_c$ limits.}.
The natural question to ask is whether this theory may be integrable.

\begin{figure}[t!!]
	\centering
	\includegraphics[width=\textwidth]{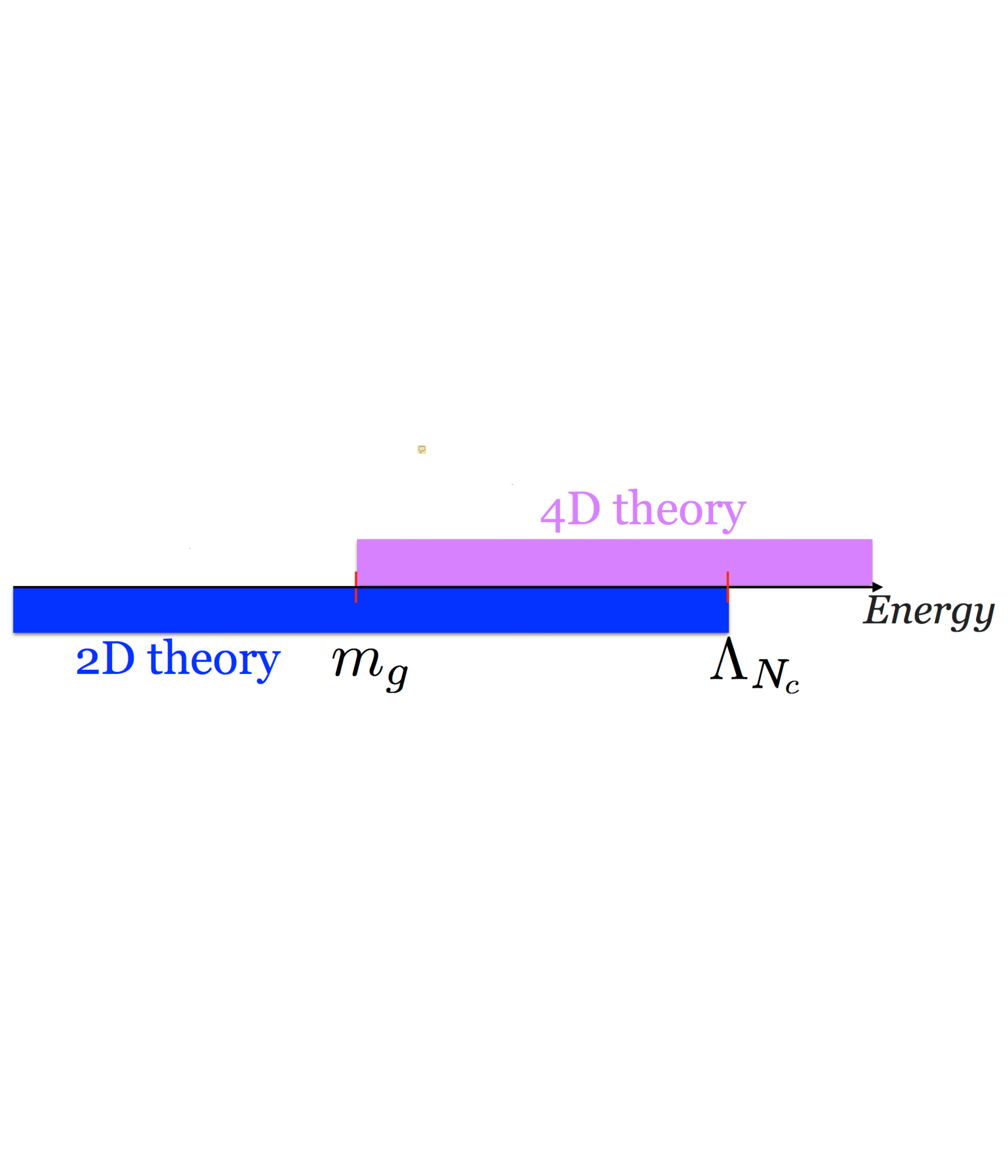}
	\caption{At large number of colors $N_c$ the worldsheet theory remains unitary up to a scale $\Lambda_{N_c}$, which is parametrically heavier than the mass of the lightest glueball.}
		\label{fig_scales}
\end{figure}

Note, that in this formulation one avoids a potential conflict between integrability, usually associated with higher spin conserved currents, and the Coleman--Mandula theorem. Indeed, in the planar limit the bulk theory describes  
non-interacting hadrons in agreement with Coleman--Mandula. This does not imply that the worldsheet theory is also free.

A more general interesting question about the worldsheet theory is what kind of asymptotic UV behavior it exhibits in the decoupling limit.
As was pointed out in \cite{Dubovsky:2012wk},  worldsheet theories of critical strings exhibit a new kind of asymptotic UV behavior (dubbed asymptotic fragility), characterized  by time delays  growing proportionally to the collision energy.  This behavior is related to a very basic geometric property of a string---an energy of a string segment is proportional to its physical length. Hence, one may expect that asymptotic fragility is a generic UV behavior of 
worldsheet theories for confining strings in the planar limit. 

These considerations open another option for integrability to play a role on the worldsheet. 
Namely, it may happen that integrability is broken softly at low energies, but gets restored in  high energy scattering on the worldsheet at large $N_c$.

The main purpose of the present paper is to take a look at these questions in the context of the
 minimal prototype theory, which is pure gluodynamics. We start in Section~\ref{sec:noint} with a brief review of universal properties of the worldsheet scattering at low energies, following \cite{Dubovsky:2012sh,Cooper:2014noa}.
Combining these with the available lattice data \cite{Teper:2009uf,Athenodorou:2010cs} (see also \cite{Juge:2002br} for the open flux tube data) immediately allows to conclude that the confining string in four-dimensional pure Yang--Mills theory cannot be integrable at $N_c=3$ and $N_c=5$. 

These considerations are based on the no-go theorem proven in \cite{Cooper:2014noa} and stating that the worldsheet integrability requires additional (on top of  translational Goldstone modes) gapless degrees of freedom on the flux tube worldsheet. Lattice data excludes extra gapless modes on the flux tube for $N_c=3,5$, hence the conclusion. However, the available lattice data does indicate the 
presence of a relatively light massive pseudoscalar state on the worldsheet \cite{Dubovsky:2013gi,Dubovsky:2014fma}. 
This state gets somewhat lighter for $N_c=5$, so more data is required to see what happens to its mass at larger $N_c$. 

The very existence of this state motivates us to explore whether it may lead to restoration of integrability at high energies.
To address this question we start in Section~\ref{sec:intext} with presenting two simple integrable extensions of the minimal worldsheet theory. The first extension is well-known and corresponds to a linear dilaton background of  fundamental strings (or, equivalently, to the light cone quantization of a non-critical string, c.f. \cite{Daszkiewicz:1997ax}). 
Another extension is characterized by the same worldsheet $S$-matrix but different symmetry properties.
Namely, an additional massless state on the worldsheet is a pseudoscalar axion in this case. Similarly to the Liouville field arising in the linear dilaton  background, the leading order operator coupled to the axion has interesting topological properties. It corresponds to the topological density which counts the self-intersection number of a string worldsheet. Interestingly, this coupling serves as good as the Liouville one (which is related to the Euler characteristic of the worldsheet) to cancel  particle production. In section~\ref{sec:general} we argue that these two examples are special cases of a larger  family of integrable models. Namely, in general one may introduce a scalar Liouville field, antisymmetric $O(D-2)$ tensor (dual to the axion) as well as symmetric traceless $O(D-2)$ tensor, with a single relation on the corresponding leading order couplings, to cancel particle production.

In Section~\ref{sec:data} we compare the value of the pseudoscalar coupling as follows from the lattice data to the special one which arises in the integrable axionic model. Quite surprisingly, we observe that the two values agree at  2.5\% percent level for $N_c=3$ if one includes statistical uncertainties only.  
The agreement persists also at $N_c=5$. This agreement is almost too good, given the quality of the lattice data and the small number of colors. 
We take a closer look at  the corresponding systematic and theoretical uncertainties and conclude that the agreement holds at $\sim 10$\% level, which is still very intriguing.

In the concluding Section~\ref{sec:last} we discuss future directions. We mainly focus on the possible steps to be made to settle whether the intriguing agreement observed in Section~\ref{sec:data} is a chance coincidence or an indication that the planar QCD string is integrable (at least at  high energies). 
\section{Integrability and extra gapless modes}
\label{sec:noint}
\subsection{No-go theorem and soft limit}
\label{sec:nogo}
Let us start with a brief review of the no-go theorem  \cite{Cooper:2014noa}, requiring the presence of additional massless states on the worldsheet, for integrability to hold.
In the absence of additional massless modes the  low energy dynamics of a long sting is described by the Nambu--Goto action,
\be
\label{braneNG}
S_{NG}=-\ell_s^{-2}\int d^{2}\sigma\sqrt{-\det\l{\eta_{\alpha\beta}+\ell_s^2\d_\alpha X^i\d_\beta X^i}\r}+\dots\;,
\ee
where $\eta_{\alpha\beta}$ is the worldsheet Minkowski metric, $X^{i}$, $i=1,\dots,D-2$, are physical transverse excitations of the string (``branons"),   $D$ is the number of space-time dimensions and $\dots$ stand for higher 
derivative terms. We are mostly interested in $D=4$ case,
however, it is useful to keep $D$ general for now. Then, the argument is based on the observation that the scattering amplitudes in this low energy effective theory are universal both at tree level and at one-loop. Indeed, out of  two possible independent one-loop counterterms, corresponding to the Einstein curvature $R$ of the induced worldsheet metric
\be
\label{indmet}
h_{\alpha\beta}=\eta_{\alpha\beta}+\ell_s^2\d_\alpha X^i\d_\beta X^i\;,
\ee
and to the rigidity term $K^2$ \cite{Polyakov:1986cs}, the former is a total derivative and the latter vanishes on-shell. 
The Nambu--Goto theory is integrable at  tree level. However,  a brute force calculation of the one-loop $2\to 4$ amplitude demonstrates that there is particle production, unless $D=26$ or $D=3$.

It is instructive to inspect the properties of a non-integrable piece of the amplitude in more details, using the Ward identities of the non-linearly realized bulk Poincar\'e symmetry as a guide\footnote{In what follows, under Poincar\'e symmetry we always understand invariance w.r.t. the full bulk Poincar\'e group, unless explicitly specified that we are talking about the two-dimensional worldsheet Poincar\'e subgroup.}.
In particular, this symmetry includes non-linearly realized boosts and rotations which act as
\be
\label{non_boost}
\updelta^{\alpha i}_\eps X^j=-\epsilon( \delta^{ij}\sigma^\alpha+X^i\d^\alpha X^j)\;,
\ee
As a consequence of this symmetry the shift current $S^i_\alpha$ of the theory is a total derivative,
\[
S^i_\alpha=\d_\alpha X^i
+\d_\beta k_\alpha^{i\beta}\;,\]
where $\d_\beta k_\alpha^{i\beta}$ is the non-linear in fields piece of the current. 
For the Nambu--Goto action it starts as 
\be
\label{cubic}
 k_{\alpha\beta}^{i}=\ell_s^2X^iT_{\alpha\beta}+\dots\;,
\ee
where $T_{\alpha\beta}$ is the energy-momentum tensor of $(D-2)$ free bosons.
Then, following the standard logic, one derives the following low energy theorem for  amplitudes with emission of a single soft branon $X^i$ of a momentum $p$
\be
    \label{double_soft}
   i \langle out|  in, p\rangle=-p_+^2\langle out|
    k_{--}^{i}(p) |in\rangle\;,
\ee
where for definiteness we consider a left-moving 
 soft branon, $p_-=0$.
Naively, this relation implies that the amplitude $\langle out|  in, p\rangle$ is double soft, {\it i.e.}, vanishes at least as the second power of the branon momentum,
\[
\langle out|  in, p\rangle\sim {\cal O}(p_+^2)\;.
\]
However, $2\to 4$ amplitudes calculated in \cite{Cooper:2014noa} do not comply with this expectation. 
This indicates the presence of  singularities on the r.h.s. of (\ref{double_soft}). These are
 related to peculiarities of Goldstone bosons in two dimensions. 

\begin{figure}[t!]
  \begin{center}
        \includegraphics[height=5cm]{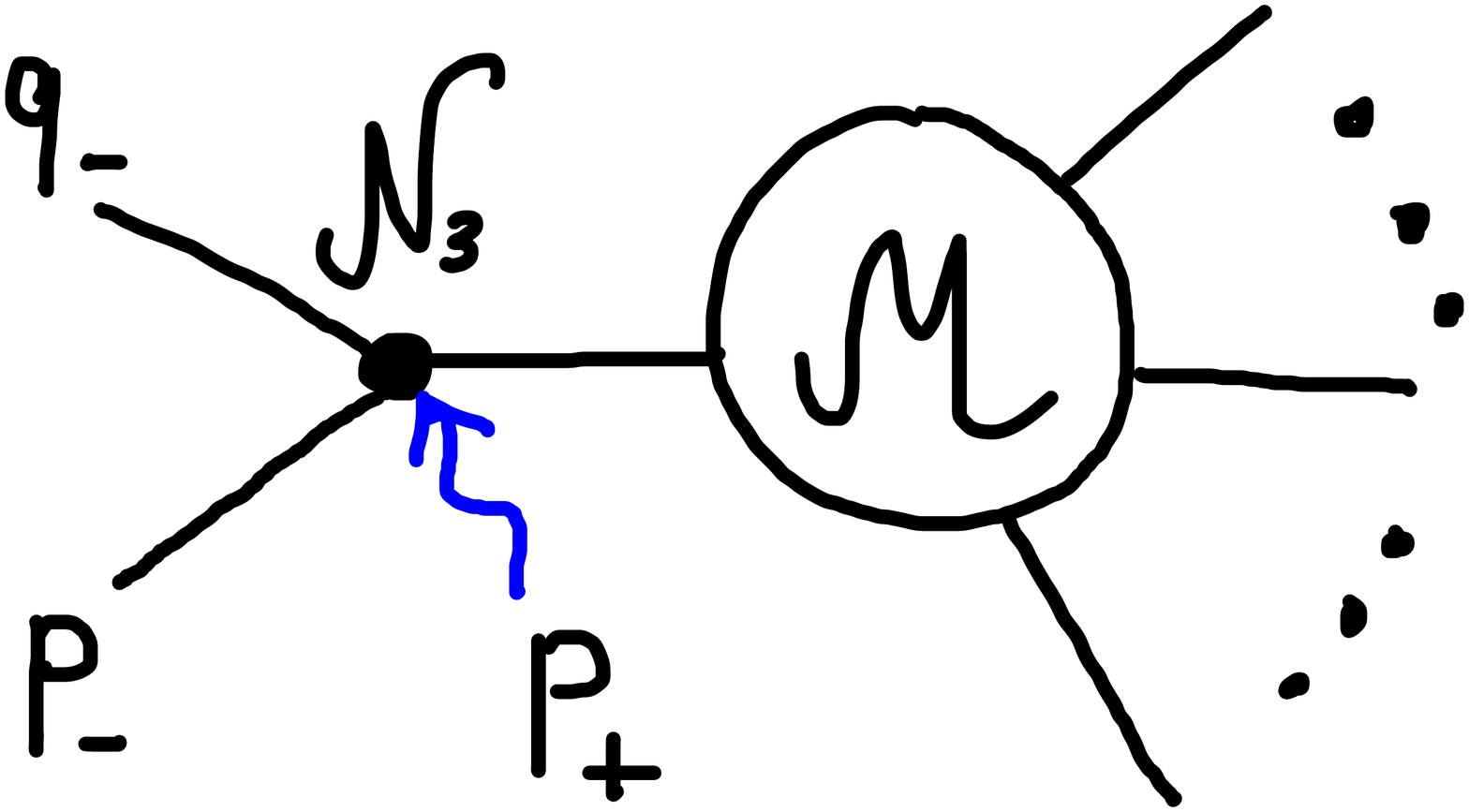}
        \hspace{1cm}
          \includegraphics[height=5cm]{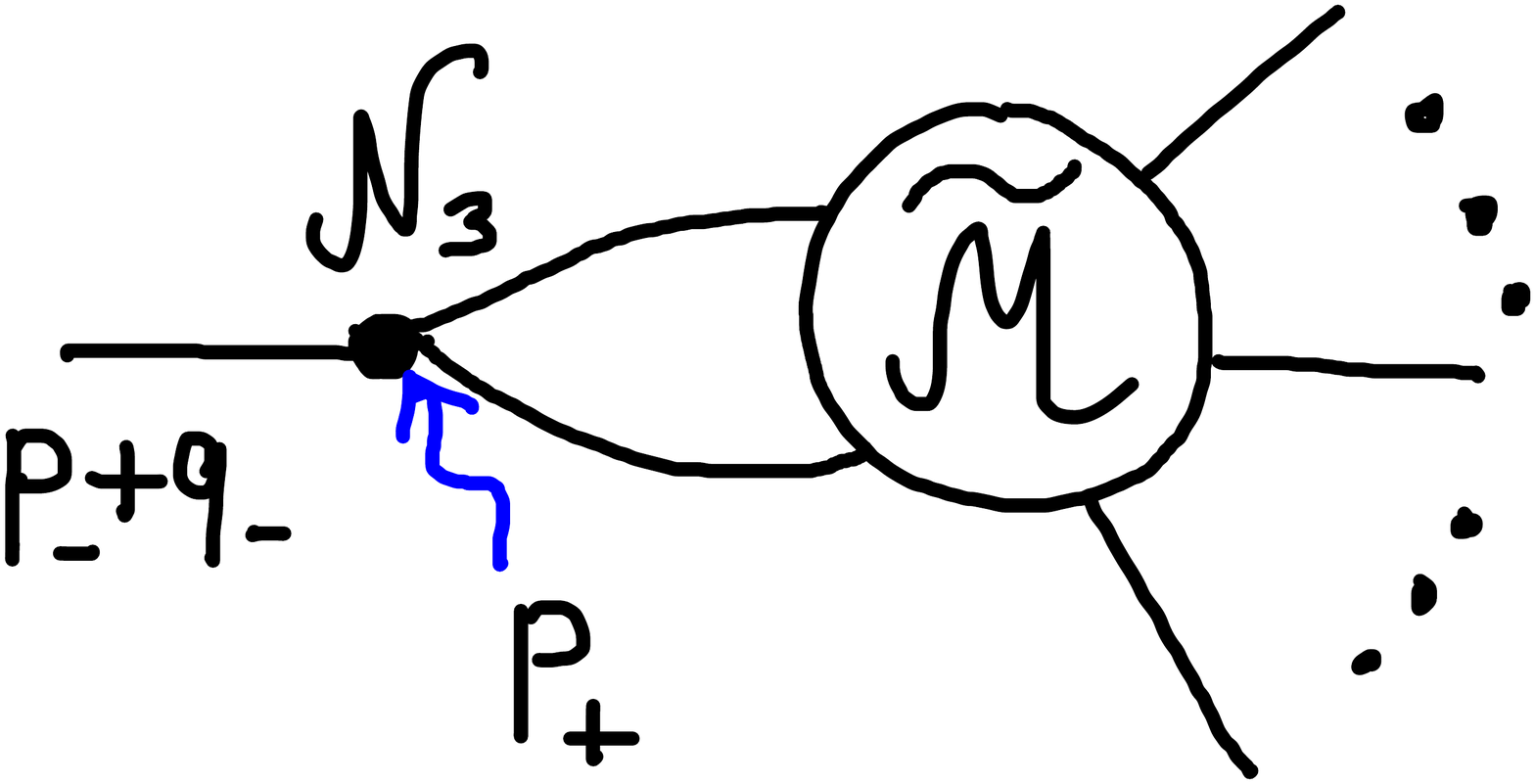}
          \\
          a)~~~~~~~~~~~~~~~~~~~~~~~~~~~~~~~~~~~~~~~~b)
            \caption{Two type of contributions giving rise to collinear singularities in the shift current Ward identities.}
        \label{fig:ward}
    \end{center}
\end{figure}
  There are two types of diagrams, which may lead to singularities.
 First, there are tree level diagrams
 of the type shown in 
Fig.~\ref{fig:ward}a). These exhibit a collinear singularity in the soft limit. Note that in higher dimensions collinear singularities do not arise in the soft limit for a generic configurations of momenta. 
In addition, there are one loop diagrams of the type shown in Fig.~\ref{fig:ward}b), which may give rise to Coleman--Thun type singularities \cite{Coleman:1978kk}, which are also possible only in two dimensions.

Note that in both cases it is only the leading cubic term in the shift current (\ref{cubic}) which gives rise to singular contributions.
Terms with larger number of legs, and/or additional derivatives necessarily involve lines with right-moving momenta and do not give rise to the violation of double softness.
\subsection{Uniqueness of $e^{i\ell_s^2s/4}$}
\label{sec:uniq}
As illustrated in \cite{Cooper:2014noa}, the relation between particle production at one loop and violation of the double softness is not only interesting from the conceptual viewpoint, but is
 also useful at the level of practical calculations. 
Namely,  Ward identities provide a less technically expensive way to calculate non-double soft pieces in the amplitude (these are also the ones, which violate integrability) than a brute force calculation of six particle amplitudes.
Let us present here a somewhat different application of the current algebra, which will prepare the base for the arguments presented in Section~\ref{sec:intext}.

Namely, the arguments above exclude integrability for $D\neq 3, 26$ in the absence of additional gapless modes.  For $D=3, 26$ we are left with a possibility of a purely reflectionless\footnote{Obviously, for a single flavor $D=3$  the scattering is always reflectionless. It is straightforward to check that for $D>4$ reflections are inconsistent with the Yang--Baxter equation for massless $O(D-2)$ invariant theories.} scattering, which in a general massless theory may be characterized by the following phase shift \cite{Zamolodchikov:1991vx},
\be
\label{Zamolodchikov}
\e^{2i\delta(s)}=\prod_j{\mu_j+s\over \mu_j-s}\e^{iP(s)}\;,
\ee
which is a general massless Castillejo--Dalitz--Dyson (CDD) \cite{Castillejo:1955ed} factor.
Here $P(s)$ is an odd polynomial in the Mandelstam variable $s$ and the CDD poles $\mu_j$ are located in the lower half of the complex plane, and come 
in pairs symmetric with respect to  the imaginary axis or belong to it.

The choice \cite{Dubovsky:2012wk}
 \be
 \label{S-matrix}
 \e^{2i\delta_{GGRT}(s)}=\e^ {i s\ell_s^2/4}\;,
 \ee
corresponds to a critical bosonic string at $D=26$ and to a light cone quantization of a bosonic string at $D=3$. In both cases these are integrable theories consistent with non-linearly realized $D$-dimensional Poincar\'e
symmetry. This leaves open the question whether any other of the CDD phase shifts (\ref{Zamolodchikov}) might also correspond to an integrable relativistic string.
Also, it would be satisfactory to see  the evidence for non-linearly realized Poincar\'e symmetry by using the exact $S$-matrix (\ref{S-matrix}) alone, without relying on the light cone quantization.
We will see now that Ward identities allow to achieve both goals. 

Let us first prove that (\ref{S-matrix}) is the only possible choice consistent with the target space Poincar\'e symmetry.
To achieve this, let us apply  the soft theorem (\ref{double_soft}) to the  amplitude of the following  $2\to 4$ process
\be
X(p_++k_+) X(p_-+q_-)\to  X(p_+)X(k_+)X(p_-)X(q_-) \;,
\ee
where for $D=26$ we chose all branons to have the same flavor, with 
 $p_+$ being a soft momentum. Integrability requires this amplitude to be zero. Of course, this implies that its non-double-soft part is also zero.
%
According to our previous discussion, violation of double softness is related to singular contributions into
\be
{\cal K}=  \langle out| k_{--}(p_+) |in\rangle\;,
    \label{kme}
\ee
which originate from the two types of diagrams presented in Figure~\ref{fig:ward}.
A tree collinear singularity, corresponding to   Figure~\ref{fig:ward}a), is given by the sum of three factorized terms
\begin{gather}
 {\cal K}_{coll} =
 { i{\cal N}_3(p_-,q_-)
 \over 2p_+
 }
 \l {i{\cal M}_4(p_-+q_-,k_+)\over p_-+q_-}-{i{\cal M}_4(p_-,k_+)\over p_-}-{i{\cal M}_4(q_-,k_+)\over q_-}\r
    \label{sing}
\end{gather}
where 
\[
{\cal N}_3(p_-,q_-)=(p_-+q_-)^2+p_- q_-
\]
is a matrix element of $k_{--}(0)$ between right-moving states with momenta $(p_-,q_-,-(p_-+q_-))$ and ${\cal M}_4(p_-,p_+)$ is a 4-particle amplitude with colliding particles carrying momenta
$p_-$ and $p_+$. The latter is related to the phase shift as 
\be
\label{M4}
i{\cal M}_4(p_+,p_-)=2s(e^{2i\delta(s)}-1)\;,
\ee
with $s=4p_+p_-$.
%

A one loop Coleman--Thun singularity can be reproduced by cutting rules and the corresponding on-shell diagram is presented in Figure~\ref{fig:on-shell}.
\begin{figure}[t!]
    \begin{center}
          \includegraphics[height=5cm]{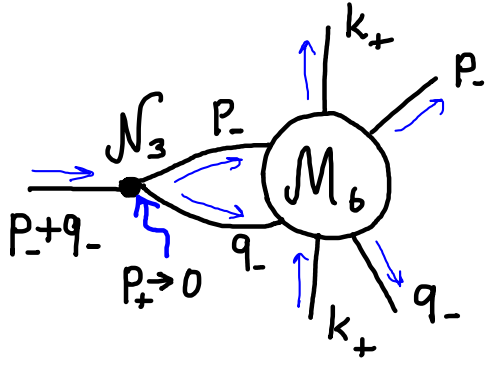}
        \caption{The on-shell diagram giving rise to the Coleman--Thun contribution into the Ward identities.}
        \label{fig:on-shell}
    \end{center}
\end{figure}
The ${\cal M}_6$ blob in this diagram represents a connected  six particle scattering amplitude. Integrability implies that it can be presented in the following factorized form
\be
i{\cal M}_6(p_-,q_-,k_+\to p_-',q_-',k_+')=i{\cal M}_4(p_-,k_+)i {\cal M}_4(q_-,k_+)\frac {2\pi \delta(p_--p_-')+2\pi \delta(p_--q_-')}{4 k_+}
\ee
Then the Coleman--Thun singularity reduces to 
\be
\label{CThun}
 {\cal K}_{CT} =
 { {\cal N}_3(p_-,q_-)i{\cal M}_4(p_-,k_+)i {\cal M}_4(q_-,k_+)\over 4k_+}{\cal I}_{sing}\;,
\ee
where ${\cal I}_{sing}$ is the singular part of the remaining loop integral
\[
{\cal I}={1\over 2\pi} \int {dl_+dl_-\l\delta(l_-)+\delta(l_--q_-+p_-)\r\over4\l(p_-+l_-)(-p_++l_+)+i\epsilon\r\l-(q_--l_-)l_++i\epsilon\r}={ i\over2 p_-p_+q_-}+reg.
\]
For non-linearly realized Poincar\'e symmetry to be consistent with the absence of particle production, the Coleman--Thun singularity  (\ref{CThun}) needs to cancel against
the collinear singularity (\ref{sing}), which requires 
\[
i\l {i{\cal M}_4(p_-+q_-,k_+)\over p_-+q_-}-{i{\cal M}_4(p_-,k_+)\over p_-}-{i{\cal M}_4(q_-,k_+)\over q_-}\r=i{i{\cal M}_4(p_-,k_+) i{\cal M}_4(q_-,k_+)\over 8k_+p_-q_-}\;.
\]
Recalling the relation (\ref{M4}) between the amplitude and the phase shift, one immediately finds that
\be
\e^{2i\delta\l 4k_+ (p_-+q_-)\r}=\e^{2i\delta\l 4k_+q_-\r}\e^{2i\delta\l 4k_+p_-\r}
\ee
hence the phase shift (\ref{S-matrix}) is the only possible one.

It is straightforward to apply the same check to amplitudes with a larger number of particles, which results in analogous relations between ${\cal M}_n$, ${\cal M}_{n+2}$ and ${\cal M}_4$,  which all hold for the $S$-matrix (\ref{S-matrix}).

%
%

Let us now present an argument proving the inverse statement, namely, that the $S$-matrix (\ref{S-matrix}) does indeed correspond to a theory with non-linearly realized target space Poincar\'e symmetry (for $D=26$ and $D=3$).
This conclusion does not follow from the previous arguments, because the corresponding current algebra proof  would require to check the Ward identities  with an arbitrary number of soft branons.

Of course, for $D=26$ this statement follows from the arguments presented in \cite{Dubovsky:2012wk}, demonstrating that (\ref{S-matrix}) describes a free critical bosonic string.
For $D=3$ one may also argue that this follows from the absence of the Poncar\'e anomaly in the light-cone quantization of $D=3$ bosonic string \cite{Mezincescu:2010yp}, and from the equivalence between (\ref{S-matrix}) and a light cone quantized string. There is a minor technical subtlety in this argument though, because light cone quantization is manifestly Poincar\'e invariant in the sector of short string, and (\ref{S-matrix}) corresponds to a long string ({\it i.e.}, a string with a unit winding in the limit when a compactification radius is taken to infinity). Related to this, arguments of \cite{Mezincescu:2010yp} do not allow to fix the normal ordering constant $a$ at $D=3$, but it does get fixed in the long string sector by demanding the worldsheet Poincar\'e symmetry \cite{Dubovsky:2014fma}. 

Apart from fixing this technicality for the $D=3$ string, the main interest in presenting a somewhat clumsy argument below is that it can be applied more generally.
In particular, for the new integrable worldsheet theory described in section~\ref{sec:ax} this  is the only currently available way to establish invariance under the target space Poincar\'e symmetry. 

The argument is based on the following more general statement. 
Let us consider an arbitrary theory of $(D-2)$ massless bosons invariant under the shift symmetry and satisfying the following properties
 \begin{enumerate}
  \item[(I)] Its $2\to 2$ scattering amplitudes agree with those in the Nambu--Goto theory both at tree level and at one loop ({\it i.e.},
  up to  the fourth and sixth orders in the derivative expansion)
  \item[(II)] The non-double soft parts of its amplitudes satisfy (\ref{double_soft}).
\end{enumerate}
Then the theory enjoys non-linearly realized $D$-dimensional Poincar\'e symmetry. 

The argument goes by induction in the number of legs $N$ and loops $L$.
  Namely, let us assume that amplitudes of the theory can be reproduced from a Lagrangian invariant under the target space Poincar\'e symmetry up to $N$ legs and $L$ loops, but fail to do so either
  at $(N+2)$ legs and $L$ loops, or at $N$ legs and $(L+1)$ loops. 
   The property (I) ensures that the failure may happen only either starting at six legs or at more than one loop for $N=4$. Note that for a shift invariant theory  a local vertex of a schematic form $\int\d^{2n}X^N$ corresponds to 
   \[
   L={N\over 2}-n
   \]
order in the loop expansion.
  
  The assumptions above imply that the action of the theory can be presented in the form
  \[
  S=S_{N,L}+V+\dots\;,
  \]
  where $S_{N,L}$ is a Poincar\'e invariant action, which agrees with our theory up to $N$ legs and $L$ loops,  $V$ is a local vertex either of the form $\int\d^{2L+2}(\d X)^N$ or $\int\d^{2L}(\d X)^{N+2}$, and $\dots$ stand for terms with larger number of legs or derivatives. Here we include in $S_{N,L}$ also all terms with larger number of legs as required by the symmetry.
  For instance, both $S_{N,0}$ and $S_{N,1}$ is the Nambu--Goto action for any $N$. 
  
  An amplitude produced by $V$ should be double soft, because a single soft part of the amplitude is fixed by lower dimensional terms in the action through  (\ref{double_soft}), and hence should be the same as in the Poincar\'e invariant theory $S_{N,L}$, which also satisfies (\ref{double_soft}).
  
  The last step in the argument is to prove that any local double soft vertex $V$ with $N>4$ or $L>1$ can be presented in the form with at least two derivatives acting on each field $X^i$ appearing in the vertex.
  Indeed, for $N>4$ double softness implies that $V$ is invariant under the Galilean symmetries
  \be
  \label{Galilean}
  X^i\to X^i+\sigma^\alpha
  \ee
  for any\footnote{A justification of this intuitively clear statement turns out to be surprisingly messy. We present the corresponding argument in the Appendix A.} $i$. 
    Then the required statement follows from the absence of non-linear multi-Galileons in two dimension \cite{Goon:2012dy} (see also Appendix A).
  
The $N=4$ case is special because in this case 
shift invariance of $V$ is enough to ensure  double softness  without requiring the Galilean invariance.
  However, it is immediate to check  that with at least eight derivatives present (which is the case for $L>1$), in two dimensions  it is always possible to rewrite  a quartic vertex $V$ in a form with at least two derivatives acting on each $X^i$.
  
  Given the expression for the extrinsic curvature,
  \[
  K^i_{\alpha\beta}=\d_\alpha \d_\beta X^i+\dots
  \]
  we see that $V$  can be written in a manifestly Poincar\'e invariant form, which completes the proof of the inductive step\footnote{In fact, if there were non-linear Galileons, one would still be able to promote $V$ into a Poincar\'e invariant term as a consequence of its Galilean symmetry, c.f. \cite{Gliozzi:2012cx}.}.
%
%
%

Note, that the above argument applies only to the part of the amplitudes, which is perturbative w.r.t. derivative expansion.
This is enough for our purposes, because we are going to apply it to (\ref{S-matrix}).
 This $S$-matrix has  all the required analyticity and unitarity properties. These guarantee  that in the regime where it can be expanded perturbatively, its expansions should be possible to match with a diagrammatic expansion following from some local Lagrangian. An additional unconventional property of this $S$-matrix is that its perturbation series in $\ell_s$ converges absolutely at all values of energy, indicating the absence of non-perturbative effects. We have already proven that at $D=3,26$ the $S$-matrix (\ref{S-matrix}) satisfies both (I) and (II), hence  our argument above indicates that it enjoys the target space Poincar\'e symmetry for these values of $D$.

It might appear surprising that we managed to establish a non-linearly realized symmetry of a theory by analyzing exclusively single-particle soft theorems, since usually the structure constants of the algebra are manifest in the two-particle soft limits \cite{PhysRevLett.16.879}.  
This extends the recent observation that the tree level Nambu--Goto action (in four dimensions) is uniquely fixed by single-particle soft limits of the amplitudes  \cite{Cheung:2014dqa}.
\subsection{Implications for gluodynamics}
Applying the no-go result of section~\ref{sec:nogo} to pure gluodynamics allows to reach some conclusions  about (non-)integrability of the QCD string. Namely, in $D=4 $ both for $SU(3)$ and $SU(5)$ groups the existence
of additional gapless degrees of freedom on the flux tube worldsheet is excluded by the existing lattice data \cite{Teper:2009uf}. This follows both from the direct analysis of flux tube excitations, and from measuring the L\"uscher correction (aka Casimir energy) to the energy of the ground state. This rules out integrability of these theories.

For $D=3$ extra massless states are not required for integrability. However, results of section~\ref{sec:uniq} indicate that the phase shift (\ref{S-matrix}) is the only one consistent with the Poincar\'e symmetry in this case.
The analysis of $SU(6)$ lattice data \cite{Athenodorou:2011rx},  presented in \cite{Dubovsky:2014fma}, revealed the presence of corrections to this phase shift at a high confidence level, thus ruling out integrability in this case as well.
 
 Of course, these results are hardly surprising.  A more interesting question is to test for integrability of pure glue in the large $N_c$ limit. This would require extending the above lattice results for gauge groups of higher rank.  In particular, the $D=4$ lattice data show the presence of a relatively light pseudoscalar state on the flux tube worldsheet  \cite{Dubovsky:2013gi}. One may wonder whether this state might become massless in the planar limit and to cancel particle production. Apart from approaching this question experimentally, {\it i.e.}, by measuring the mass of this state at large $N_c$ on a lattice, this possibility may be also assessed from a theoretical side. Namely, the question is what options are available for constructing  integrable worldsheet theories in the presence of additional massless modes, and in particular whether a single additional massless pseudoscalar state may do the job. 
 
\section{Simple integrable extensions of the minimal string}
\label{sec:intext}
A large set of candidates for a massless sector of an integrable worldsheet theory may be constructed following the close connection between integrability and critical strings. Indeed, as we already saw, requiring integrability  for  the  minimal Nambu--Goto theory allows to identify the critical bosonic string, $D=26$ (modulo a somewhat degenerate $D=3$ case). The same pattern repeats for superstrings in the Green--Schwartz description \cite{Cooper:2014noa}. A tree level integrability singles out classical $\kappa$-symmetric Green--Schwartz superstrings, and at one loop we are  left with $D=10$ and $D=3$ options.

This set of examples can be naturally expanded in the following way. Let us consider a compactification of the critical bosonic string (a generalization to critical superstrings is straightforward), which can be described by the Polyakov 
worldsheet action of the form,
\be
\label{critical}
S=\int d^2\sigma \sqrt{-g}\l-{1\over 2}\l\d_\alpha X^\mu\r^2+{\cal L}_{CFT}(g_{\alpha\beta},\phi^a)\r\;,
\ee
where $\mu=0,\dots,3$ and ${\cal L}_{CFT}$ is a Lagrangian of a $c=22$ conformal field theory (CFT).
Then the worldsheet theory,  describing perturbations around a long string ($X^0=\tau$, $X^1=\sigma$) embedded into this background, will be integrable.
Note that for a general interacting CFT the worldsheet $S$-matrix is not well-defined because of the IR divergences.
However, even if so, scattering amplitudes of branons will be IR safe and described by the  integrable $S$-matrix 
(\ref{S-matrix}). This can be checked by, e.g., fixing the conformal gauge and performing the light cone quantization, which is compatible with the target space Poincar\'e symmetry
for a critical string. The resulting  spectrum is a generalization of the flat space spectrum, described in \cite{Caselle:2013dra}. At the level of scattering amplitudes (if these can be defined), this procedure corresponds to gravitational dressing introduced in \cite{Dubovsky:2013ira}.
Interestingly, an example of a gauge theory with a massless spectrum on the flux tube worldsheet agreeing with a critical superstring for a certain background was suggested recently in \cite{Shifman:2015kla} (although, with a somewhat different motivation). 

One may be skeptical though about this set of examples, because critical strings contain massless graviton in the spectrum of short strings, which cannot appear in the spectrum of  a quantum field theory \cite{Weinberg:1980kq}.
However, it is not obvious that integrability and the critical spectrum of massless states on a long string are enough to guarantee that the spectrum of short strings is also the same as for the critical string. In particular,
it might be possible to add massive modes on the  worldsheet in a way which preserves integrability and changes the spectrum of short strings\footnote{Although, our preliminary results strongly suggest that introducing massive fields on the worldsheet breaks integrability, at least in the perturbative regime.}.

It is interesting to understand whether integrability necessarily implies that the spectrum of massless modes on a flux tube coincides with that for some critical string.
$D=3$ bosonic and super- strings mentioned above already suggest that this is not
necessarily the case, although these counterexamples may appear somewhat degenerate. The main goal of this section is to extend this set of examples. Interestingly, these include  a string in four dimensions, and with the same matter content as on the worldsheet of the QCD flux tube. Namely, the only additional particle on the worldsheet will be a pseudoscalar. At the moment we are unaware how to fit this  example and its generalizations described below
in the conventional string theory framework.

Given that currently we can only construct this theory using the techniques described in the previous section, we find it instructive to illustrate first 
how these work in a more familiar case, which allows for the Polyakov description (\ref{critical}).
 For this purpose we consider perhaps the simplest choice of ${\cal L}_{CFT}$ in (\ref{critical})---the linear dilaton theory\footnote{Note that a linear dilaton background
is often referred to as a non-critical string. Instead, we mostly use the terminology, where we call critical any worldsheet theory of the form (\ref{critical}) with  total central charge equal to 26.}.
\subsection{Adding a worldsheet scalar (linear dilaton)}
\label{sec:dil}
The linear dilaton CFT is characterized by the following action
\be
\label{lindil}
S_{ld}=\int d^2\sigma\sqrt{-g}\l {1\over 2}\l\d \phi\r^2+Q R\phi \r\;,
\ee
where $R$ is the Einstein curvature of the Polyakov metric $g_{\alpha\beta}$.  The central charge of this CFT is equal to 
\be
\label{cld}
c_\phi=1+{48 \pi Q^2}\;,
\ee
so that for the worldsheet theory (\ref{critical}) to be critical one needs
\[
Q=\sqrt{25-D\over 48\pi}=\sqrt{7\over 16\pi}\approx 0.373176\dots\;,
\]
where we plugged in the $D=4$ value.

It is straightforward to guess what is the corresponding integrable $S$-matrix on the worldsheet of a long string in the linear dilaton background. This is an integrable theory describing $(D-2)$ massless branons and an additional
massless scalar dilaton. At the lowest order in the derivative expansion it  coincides with the Nambu--Goto theory describing a string propagating in $(D+1)$ dimensional target spacetime. So it is natural to expect that it should be
described by the $S$-matrix (\ref{S-matrix}) with $(D-1)$ flavor. This expectation is confirmed by the observation \cite{Daszkiewicz:1997ax} that the spectrum of short strings in the linear dilaton background agrees with the spectrum  obtained by the light cone quantization of a bosonic string in a non-critical dimension.

Let us see now how to arrive at this result using the methods of section~\ref{sec:nogo}. We need to show that even though the $S$-matrix (\ref{S-matrix}) with $(D-1)$ flavors is not compatible with a non-linearly realized $ISO(D,1)$ symmetry for 
$D+1\neq 3,26$, it still enjoys a non-linearly realized  $ISO(D-1,1)$ symmetry. We will follow the logic of section~\ref{sec:uniq} and show that one can construct a shift invariant action depending on  fields
\[
Y^A \equiv (X^i,\phi)\;,
\]
$i=1,\dots,D-1$, which is invariant under $ISO(D-1,1)$ and reproduces (\ref{S-matrix}) order by order in the number of legs $N$ and in the number of loops $L$ (as before, the loop expansion corresponds to the derivative expansion).
It is instructive to see explicitly what happens at the first few orders.
At the lowest derivative level one starts with the $(D+1)$-dimensional Nambu--Goto action,
\be
\label{S0}
S_{0}=-\ell_s^{-2}\int d^{2}\sigma\sqrt{-\det\l{h_{\alpha\beta}+\ell_s^2\d_\alpha \phi\d_\beta\phi}\r}\;,
\ee
where the induced metric on the worldsheet $h_{\alpha\beta}$ is given by (\ref{indmet}), as before. 
As we already know, this action agrees with the $S$-matrix (\ref{S-matrix}) at the tree level, {\it i.e.} for $L=0$ and any $N$.
For one loop two-to-two scattering this action gives rise to the Polchinski--Strominger annihilation vertex \cite{Polchinski:1991ax},\cite{Dubovsky:2012sh}, which has to be canceled to match (\ref{S-matrix}). In principle,
this can be achieved by introducing the following local term in the action
\be
\label{PS}
S_{PS}={D-25\over192\pi}\ell_s^4\int d^2\sigma \d_\alpha\d_\beta Y^A\d^\alpha\d^\beta Y^A\d_\gamma Y^B\d^\gamma Y^B\;.
\ee
Unfortunately, this term is inconsistent even with the restricted $ISO(D-1,1)$ Poincar\'e symmetry acting only on $X^i$'s. However, the linear dilaton action (\ref{lindil}) suggests 
another path for canceling annihilations. Namely, let us add the linear dilaton coupling of the from
\be
\label{S1}
S_{ld}=Q_d\int d^2\sigma \sqrt{-h}\phi R_h\;,
\ee
where $R_h$ is the scalar curvature of the induced worldsheet metric. 
This vertex does not break the shift symmetry of the Liouville field $\phi$ because
\[
\chi=2-2g={1\over 4\pi}\int d^2\sigma \sqrt{-h} R_h
\]
is a topological invariant---Euler number---of the worldsheet.
Naively, the vertex (\ref{S1}) is lower order than (\ref{PS}) (it starts with $N=3$ and four derivatives). However,  (\ref{S1})  vanishes on-shell and can in principle be shifted into a higher order term by performing a field redefinition. However, we prefer to keep intact the transformation law (\ref{non_boost}) and to leave (\ref{S1}) as it is. Note that the  transformation of the Liouville field $\phi$ under non-linearly realized boosts is 
\[
\updelta^{\alpha i}_\eps \phi=-\epsilon X^i\d^\alpha \phi\;.
\]
The fastest way to calculate the contribution of (\ref{S1}) into scattering amplitudes, though,  is to make use of the field redefinition. Namely, (\ref{S1}) expanded up to cubic order takes the following form,
\be
\label{S1exp}
S_{ld}={Q_d\ell_s^2}\int d^2\sigma\phi\l
\d_\alpha\l \d^\alpha X^i\d^2X^i\r-{1\over 2}\d^2(\d X^i)^2+\dots
\r\;.
\ee
This vertex can be removed by the following field redefinition,
\begin{gather}
\phi\to \phi +{Q_d\ell_s^2\over 2}(\d X^i)^2\\
X^i\to X^i+Q_d\ell_s^2\d_\alpha\phi\d^\alpha X^i\;,
\end{gather}
which instead results in the following quartic vertex
\be
\tilde{S}_{ld}=Q_d^2\ell_s^4\int d^2\sigma\l-{1\over 8}\l\d_\alpha(\d X^i)^2\r^2-{1\over 2}\l\d_\beta\l \d_\alpha\phi\d^\alpha X^i\r\r^2\r\;,
\ee
which agrees on-shell with the Polchinski--Strominger one (\ref{PS}) if one sets
\[
Q_d=Q\;.
\]
This way we constructed a Poincar\'e invariant action which reproduces the phase shift  (\ref{S-matrix}) up to $N=4$, $L=1$.
The rest of the argument proceeds in a close analogy to section~\ref{sec:uniq}. Indeed, the Ward identity (\ref{double_soft})
remains valid. The shift current still can be presented in the form (\ref{cubic}) with $T_{\alpha\beta}$ 
being the energy-momentum tensor of $(D-1)$ free bosons with an additional linear dilaton contribution $T^{ld}_{\alpha\beta}$ originating from (\ref{S1exp})
\be
\label{Tld}
T^{ld}_{\alpha\beta}=-2Q\ell_s^2\d_\alpha\d_\beta\phi\;.
\ee
The corresponding bilinear terms in the current will in general also give rise to singular contribution on the r.h.s. of (\ref{double_soft}). This type of singularities is not special to two dimensions, similar singularities arise also in pion-nucleon scattering (see, e.g., \cite{Bando:1987br} for a review). However, these terms will not affect our inductive argument, due to the special structure of the $S$-matrix (\ref{S-matrix}).
Indeed, in general this contribution controls the emission of soft branons from external legs through diagrams of the type presented  in Figure~\ref{fig:softld}. 
\begin{figure}[t!]
    \begin{center}
          \includegraphics[height=5cm]{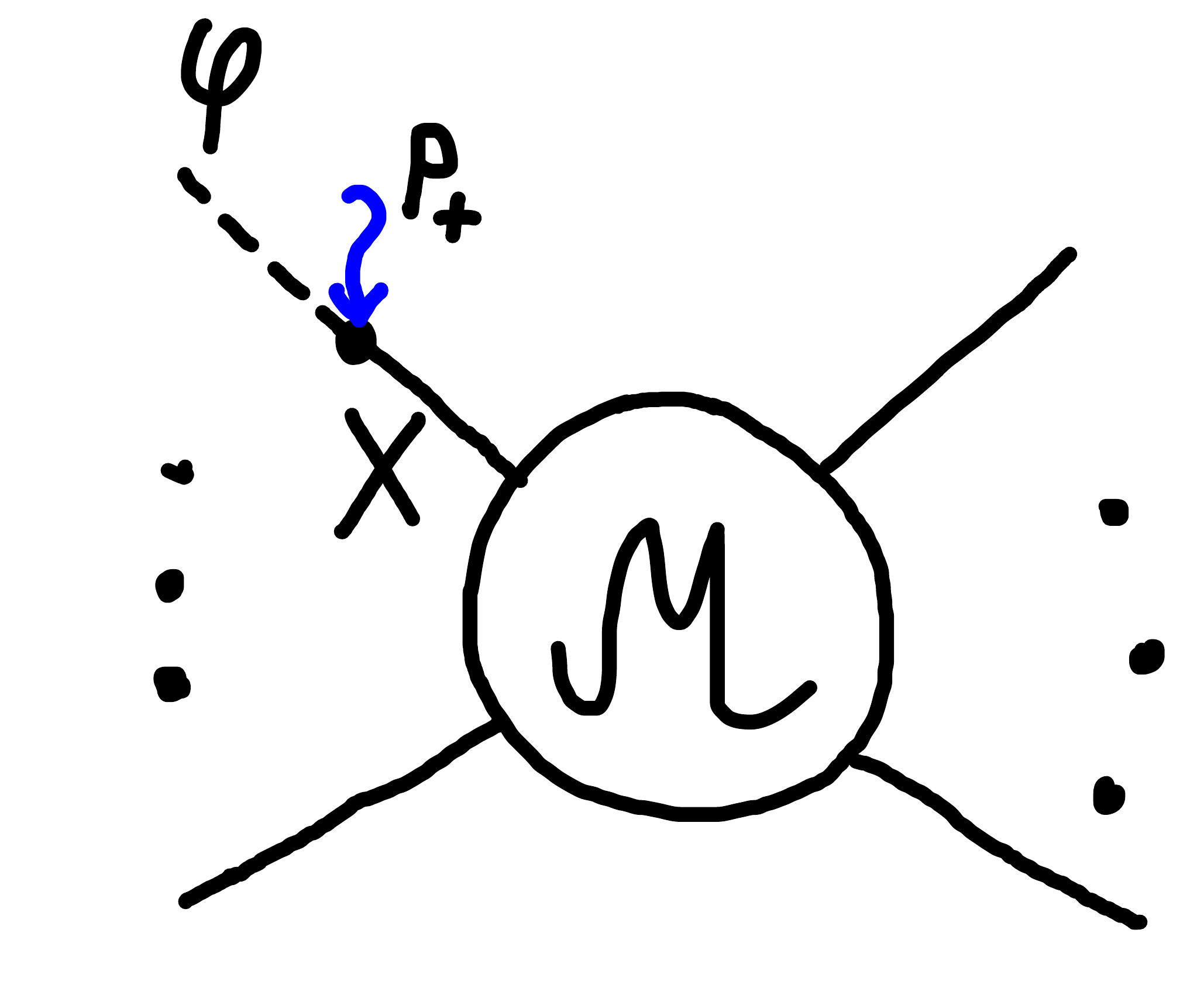}
        \caption{A bilinear contribution  into the Ward identities.}
        \label{fig:softld}
    \end{center}
\end{figure}
 At every step in the inductive argument all lower order amplitudes agree
with the expansion of (\ref{S-matrix}) and then these kind of diagrams cancel out. Clearly, this cancellation is necessary for the integrable theory to be constructed. Let us illustrate  how the cancellations work using the following 
process as an example,
\[
X(p_++k_+) \phi(p_-)\to  X(p_+)X(k_+)X(p_-) \;,
\]
where for simplicity we restrict to branons of the same flavor.
The soft theorem (\ref{double_soft}) relates the corresponding amplitude ${\cal M}_5$ to the matrix element,
\[
i{\cal M}_5=2Q\ell_s^2p_+^2\langle X(k_+)X(p_-)|X\d_-^2\phi (p_+)|X(k_+) \phi(p_-)\rangle+\dots\;,
\]
where dots stands for non-singular contributions. The possible single soft part of ${\cal M}_5$ takes then the following form
\be
\label{Xbrem}
i{\cal M}_5=2Q\ell_s^2p_+p_-^2\l {{\cal M}_4\over p_-}-{\tilde{{\cal M}}_4\over p_-}\r
\ee
where ${\cal M}_4$ and $\tilde{{\cal M}}_4$ are the amplitudes for
\[
X(k_+) \phi(p_-)\to  X(k_+)\phi(p_-)
\]
and
\[
X(k_+) X(p_-)\to   X(k_+)X(p_-)\;.
\]
For $S$-matrix (\ref{S-matrix}) these two amplitudes are equal to each other and (\ref{Xbrem}) vanishes as required for integrability.
Clearly, these cancelations persist in a similar manner for more complicated processes as well. Note, that even though the linear dilaton theory is not invariant with respect to rotations between $X^i$ and $\phi$, this argument demonstrates that integrability requires the phase shift to be universal, {\it i.e.} to be independent of whether the scattering particles are branons  or $\phi$.

From this point on the argument goes exactly in the same way as in section~\ref{sec:uniq}. The only minor difference is that a local vertex $V$, which has to be added at every step to match the expansion of (\ref{S-matrix}), is not expected to be double soft w.r.t. to the momentum of the Liouville field $\phi$. However this is not an obstacle to present it in $ISO(D-1,1)$ invariant form as well.

\subsection{A new integrable string theory from a worldsheet axion}
\label{sec:ax}
In the previous section we presented what may appear a somewhat baroque way to arrive at the linear dilaton background. Definitely, for linear dilaton the conventional Polyakov approach \cite{Polyakov:1981rd} provides a much more efficient way to define the theory. However, the linear dilaton setup  does not appear to be relevant to the QCD string anyway. The only additional (massive) particle revealed in the lattice data is a pseudoscalar worldsheet axion rather than a scalar Liouville-like mode. Interestingly, as we show now, the current algebra techniques allow to identify another integrable worldsheet model. In four dimensions particle production in this model is cancelled by an extra {\it pseudoscalar} mode. This possibility is related to the existence of the following topological invariant on a  worldsheet of a string propagating in a four-dimensional space-time \cite{Polyakov:1986cs},
\be
\label{selfintersect}
\nu={1\over 8\pi}\int d^2\sigma K^i_{\alpha\gamma}K^{j\gamma}_\beta\epsilon^{\alpha\beta}\epsilon_{ij}
={1\over 32\pi}\int d^2\sigma\sqrt{-h}h^{\alpha\beta}\epsilon_{\mu\nu\lambda\rho}\d_\alpha t^{\mu\nu}\d_\beta t^{\lambda\rho}
\;,
\ee
where 
\[
t^{\mu\nu}={\epsilon^{\alpha\beta}\over\sqrt{-h}}\d_\alpha X^\mu\d_\beta X^\nu\;,
\]
and $X^\mu$ are four-dimensional embedding functions of the string worldsheet,
\[
X^\mu=\l\sigma^\alpha, X^i\r\;.
\]

Geometrically, $\nu$ counts the (sign weighted) self-intersection number of the string worldsheet.  A similar invariant can always be defined given an embedding of a $2r$-dimensional surface into a $4r$-dimensional target space (for a half-integer $r$ it is defined only mod 2). 
It was suggested that this invariant is related to the QCD $\theta$-term \cite{Mazur:1986nr}. The existence of this invariant suggests that for a four dimensional string one may replace the linear dilaton coupling (\ref{S1}) with the corresponding axionic coupling
\be
\label{S1a}
S_a={Q_a\over 4}\int d^2\sigma \sqrt{-h}h^{\alpha\beta}\epsilon_{\mu\nu\lambda\rho}\d_\alpha t^{\mu\nu}\d_\beta t^{\lambda\rho}a
\ee
to cancel particle production on the worldsheet.
Similarly to the dilaton coupling (\ref{S1}) the axion coupling formally appears to be lower order than the Polchinski--Strominger vertex, however, it also vanishes on-shell. The cubic coupling following from (\ref{S1a}) takes the following form
\be
\label{S1aexp}
S_{a}=Q_a\ell_s^2\int d^2\sigma a \epsilon_{ij}\epsilon^{\alpha\beta}\d_\alpha\d_\gamma X^i\d_\beta\d^\gamma X^j+\dots\;.
\ee
The field redefinition required to remove (\ref{S1aexp}) is
\begin{gather}
a \to a-{Q_a\ell_s^2\over 2}\epsilon_{ij}\epsilon^{\alpha\beta}\d_\alpha X^i\d_\beta X^j\\
X^j\to X^j-Q_a\ell_s^2\epsilon^{\alpha\beta}\epsilon_{ij}\d_\beta a\d_\alpha X^i\;.
\end{gather}
If we now set
\[
Q_a=Q
\]
then the resulting quartic vertex takes on-shell the following form
\be
\label{extra}
\tilde{S}_{a}=S_{PS}-{Q^2\ell_s^4\over 4}\int d^2\sigma\l\d a \r^2\l\d_\alpha\d_\beta X^i\r^2\;.
\ee
The last term in (\ref{extra}) is invariant under the Galiean shifts of $X^i$, so that it can be promoted to a Poincar\'e invariant term and subtracted from the action. 
So we conclude that the axionic theory  is as successful in canceling the one-loop Polchinski--Strominger amplitude as the diatonic one.

The contribution of (\ref{S1aexp}) into the shift current  is still a total derivative, as required by the Poincar\'e symmetry. However, the explicit form of the corresponding non-linear piece $k^i_{\alpha\beta}$ is different from (\ref{cubic}). Instead, the axionic coupling (\ref{S1aexp}) contributes to $k^i_{\alpha\beta}$ as
\be
\label{kax}
Q\ell_s^2 \epsilon_{ij}X^j(\epsilon_{\alpha\gamma}\d^\gamma\d_\beta a+\epsilon_{\beta\gamma}\d^\gamma\d_\alpha a)\;.
\ee
This strongly suggests that the axionic mechanism for canceling particle production has a fundamentally different geometric origin from the linear dilaton case. At the technical level the form (\ref{cubic}) comes out whenever all dependence of the action on the embedding functions $X^i$ of the string is through the induced metric 
(\ref{indmet}). As a result, shifts of $X^i$'s are equivalent to variation of the action w.r.t. to the flat worldsheet metric $\eta_{\alpha\beta}$, which gives rise to (\ref{cubic}).
The self-intersection number characterizes the extrinsic geometry of the string worldsheet and cannot be written as a functional of the induced metric alone. The shift current takes then a very different form. 

Related to this, it is challenging to rewrite the axion coupling
(\ref{S1a})  in the Polyakov formalism in a useful way. Of course, one can switch to the Polyakov description by replacing the induced metric $h_{\alpha\beta}$ in (\ref{S1a}) with  
the independent Polyakov metric $g_{\alpha\beta}$ (see \cite{Hellerman:2014cba} for a detailed discussion). However, this still leaves us with higher dimensional non-linear coupling between $X$'s and $\phi$, so it is not clear whether one gains much from doing that.

In spite of these differences, the current algebra argument proceeds in a complete analogy to the linear dilaton case. Namely, just like in the linear dilaton case, singular contributions of the bilinear 
terms in the shift current corresponding to (\ref{kax}) cancel out when lower order amplitudes agree with  (\ref{S-matrix}). The rest of the argument proceeds as in section \ref{sec:uniq}, and demonstrates that one may start
with (\ref{S0}) and (\ref{S1a}) and construct order-by-order in the derivative expansion a Poincar\'e invariant theory reproducing the $S$-matrix (\ref{S-matrix}).

It may appear somewhat surprising that we found two physically distinct theories (four-dimensional relativistic string with a linear dilaton or with an axion), described by the same $S$-matrix.
However,  quantum numbers of  asymptotic states in the two theories are different, so the theories are not equivalent.

Hopefully, just like for conventional critical strings, eventually it will be possible to find a more convenient formalism for building the theory than a brute force perturbative construction presented here.
However, the message of this construction appears to be rather transparent. The only obstruction for building an integrable Poincar\'e invariant theory arises at one-loop order. If there is a way to cancel 
particle production at one-loop, it can be achieved at higher loop orders as well. This fits well with many other situations when it is enough to ensure anomaly cancellation at one-loop order. In the present case the anomaly arises due to a conflict between a non-linearly realized Poincar\'e symmetry and an infinite-dimensional symmetry responsible for the classical integrability of the Nambu--Goto theory.
\subsection{Towards a general picture}
\label{sec:general}
We see  that in four dimensions it is possible to cancel the Polchinski--Strominger annihilation by adding a new particle either in the scalar or in the antisymmetric tensor (pseudoscalar) channel. It is natural to ask whether one may achieve cancellations also by adding a particle in the symmetric tensor channel and whether cancelations due to antisymmetric tensor exchange can be extended to strings propagating in higher dimension, just like it happens for the linear dilaton case. These questions are somewhat beyond the main scope of this paper, which is mainly concerned with four-dimensional strings, which may be relevant for large $N_c$ limit of confining gauge theories (in particular, for pure gluodynamics). So we will present a preliminary argument suggesting that the answer to both questions is affirmative, and leave the details for a separate dedicated study.

Following the same strategy as above, in general number of dimensions it is natural to introduce either $O(D-2)$ antisymmetric  $a_{ij}$ or symmetric traceless $s_{ij}$ tensor fields (which are scalars on the worldsheet) and to use couplings of the form
\be
\label{S1aD}
S_a={Q_a}\int d^2\sigma a_{ij}K^i_{\alpha\gamma}K^{j\gamma}_\beta\epsilon^{\alpha\beta}
\ee
or
\be
\label{S1sD}
S_s={Q_s}\int d^2\sigma \sqrt{-h}s_{ij}K^i_{\alpha\gamma}K^{j\gamma\alpha}
\ee
to cancel the Polchinski--Strominger interaction. Let us check that couplings (\ref{S1aD}) and ({\ref{S1sD})  have very similar properties to the couplings (\ref{S1}), (\ref{S1a}) which we used before. Namely, they may be presented in a fully covariant form, do not break the shift symmetry of $a_{ij}$ and $s_{ij}$, and vanish on-shell. To demonstrate this, let us consider the following total derivative,
\be
\label{totalI}
I^{\mu\nu}_{\beta\gamma}\equiv\nabla_\alpha\l\d_\beta X^\mu\nabla^\alpha\d_\gamma X^\nu-\d^\alpha X^\nu\nabla_\beta\d_\gamma X^\mu\r\;,
\ee
and extend $a_{ij}$, $s_{ij}$ to full $D$-dimensional tensors $a_{\mu\nu}$, $s_{\mu\nu}$ subject to covariant constraints
\[
a_{\mu\nu}\d_\alpha X^\nu=s_{\mu\nu}\d_\alpha X^\nu=0\;.
\]
Then, by making use of the expression
\[
K^\mu_{\alpha\beta}=\nabla_\alpha\d_\beta X^\mu
\]
for the extrinsic curvature,
 the couplings (\ref{S1}), (\ref{S1aD}) and (\ref{S1sD}) can be written as
\begin{gather}
S_{ld}=Q_d\int d^2\sigma\sqrt{-h}\phi I^{\mu\alpha}_{\mu\alpha}
\label{Ild}\\
S_{a}=Q_a\int d^2\sigma a_{\mu\nu}\epsilon^{\alpha\beta}I^{\mu\nu}_{\alpha\beta}
\label{Ia}
\\
S_{s}=Q_s\int d^2\sigma \sqrt{-h}s_{\mu\nu}I^{\mu\nu\alpha}_{\alpha}\;,
\label{Is}
\end{gather}
where to see the equivalence between (\ref{Ild}) and (\ref{S1}) one needs to make use of the Gauss--Codazzi equality.
This representation makes it manifest that none of these couplings breaks the shift symmetry of $\phi$, $a_{\mu\nu}$ and $s_{\mu\nu}$. Also, it is straightforward to check that all the couplings vanish on-shell, so one can trade them for quartic vertices as before.
Given that the only non-trivial six derivative quartic vertex for $X^i$'s is the Polchinski--Strominger one, with an appropriate choice of $Q_a$ or $Q_s$ we will be able to cancel the annihilation term for  $X^i$'s as before. Just as in section~\ref{sec:ax} it should be possible then to cancel the remaining annihilations between $X^i$'s and $a_{\mu\nu}$ or $s_{\mu\nu}$ with Poincar\'e invariant local terms and to run the inductive argument.

Notice, that the above reasoning applies also in the situation when all three couplings (\ref{Ild}), (\ref{Ia}) and (\ref{Is}) are present simultaneously and strongly suggest the existence of a larger family of integrable Poincar\'e invariant models.
The condition on the coupling constants imposed by integrability in the case of $n_d$ dilatons, $n_a$ axions and and $n_s$ tensors reads
\be
D+n_d+\frac{(D-2)(D-3)}{2} n_a+\frac{(D-1)(D-2)-2}{2} n_s+48 \pi \sum\l Q_d^2+ Q_a^2- Q_s^2\r=26.
\ee
Note that $Q_s$ contributes with the sign opposite to all other terms and consequently allows to extend the family of theories to $D>26$ which is not possible for conventional non-critical strings.
We leave for the future a detailed study of this family of models and are turning back now to $D=4$ axionic strings and  gluodynamics.

\section{Back to pure glue: a hint of integrability }
\label{sec:data}
Worldsheet dynamics of confining strings in gauge theories can be explored using lattice simulations. Pure gluodynamics is the most studied theory, and the most recent results on the spectrum of closed strings in $SU(3)$ and $SU(5)$ gluodynamics
in $D=4$ dimensions are presented in \cite{Athenodorou:2010cs} (older open string data can be found in \cite{Juge:2002br}, and
$D=3$ data can be found in \cite{Athenodorou:2011rx,Athenodorou:2013ioa}). Up until recently theoretical interpretation of this data was problematic, because the conventional perturbation theory for calculating the effective string spectra \cite{Luscher:1980ac,Luscher:2004ib,Aharony:2010db,Aharony:2013ipa} produces badly divergent asymptotic series for the energies of excited strings states in the parameter range accessible for current lattice simulations. 

The situation changed recently, after an alternative perturbative scheme based on Thermodynamic Bethe
Ansatz (TBA) has been developed \cite{Dubovsky:2013gi,Dubovsky:2014fma}. 
In particular, this resulted in  identification of a massive {\it pseudoscalar} excitation on the worldsheet of confining flux tubes in $D=4$ gluodynamics---the worldsheet axion. This excitation is present both in $SU(3)$ and $SU(5)$ data, as well as in the open string spectra.
The leading operator describing interactions of the worldsheet axion with branons is the topological coupling (\ref{S1a}), so it is natural
to compare the value of the corresponding coupling $Q_a$ as determined from the lattice data, to the special value 
\be
\label{QI}
Q\approx 0.373\dots\;,
\ee
discussed in section~\ref{sec:ax}. Remarkably, the lattice value quoted in \cite{Dubovsky:2013gi,Dubovsky:2014fma} reads{\footnote{The coupling $\alpha\approx 9.6\pm0.1$ of \cite{Dubovsky:2013gi,Dubovsky:2014fma} is related to $Q_a$ as $Q_a=\alpha/(8\pi)$.}},
\be
\label{QL}
Q_L\approx 0.382\pm0.004\;,
\ee
{\it i.e.}, it agrees with $Q$ at  $2.5\%$ level. The uncertainty quoted in (\ref{QL}) is statistical only and  the authors of  \cite{Dubovsky:2013gi,Dubovsky:2014fma} did not mean to provide a high precision measurement of $Q_a$,
for sure not at $2.5\%$ level.
This surprising coincidence definitely warrants further investigation. The main goal of the present section is to assess systematic and theoretical uncertainties entering into determination of $Q_a$ from the current lattice data.
The conclusion is that the coincidence is indeed present, but to be on a conservative side we would not claim it to hold with better than $10\%$ precision. In  section~\ref{sec:last} we will comment on possible interpretations of this intriguing coincidence, and discuss how it can be checked/sharpened with the future lattice data.

Let us start with a brief summary of the TBA technique which allows to identify the worldsheet axion and to determine its parameters (the mass $m$ and the coupling $Q_a$).
A detailed review of the TBA method can be found in  \cite{Dubovsky:2014fma}. 

Lattice simulations allow to determine energies of various excited string states at different values of the flux tube length $R$. 
In our quantitative analysis we will only use two particle string excitations with a single left-moving and a single right-moving branon propagating along the worldsheet with minimal values of the Kaluza--Klein momentum.
Higher excited states were used in \cite{Dubovsky:2014fma} for qualitative consistency checks, however the corresponding error bars are too large to use them in the quantitative analysis.
We also use the high precision ground state data which allows to fix the string tension $\ell_s$.

There are three distinct two particle states, classified according to their $O(2)$ quantum numbers---the scalar $0^{++}$,  the pseudoscalar $0^{--}$ and the traceless symmetric tensor states (the latter come in two
 polarizations $2^{++}$ and $2^{+-}$).
The energies of these states are presented by blue, red and green set of points in Figures~\ref{fig:plots3}, \ref{fig:plots5} for $SU(3)$ and $SU(5)$ gauge groups. Note that in these plots we subtracted the dominant linearly growing classical contribution due to string tension. For comparison, in Figure \ref{fig:plotsC} we presented also the total energy to give an idea of the size of the effects that we are discussing. As can be seen from the plots, spin 2 polarizations $2^{++}$ and $2^{+-}$ have different energies. This is a lattice artifact
resulting from breakdown of the $O(2)$ transverse rotational symmetry to its ${\mathbb Z}_4$ subgroup on a square lattice. It is argued in  \cite{Athenodorou:2010cs} that this effect
is larger than discretization effects in other channels due to a different scaling with the lattice spacing. It presents the major source of systematic uncertainties for our analysis.

The key idea of the TBA method is to relate the finite volume spectrum with the infinite volume phase shift for the scattering of $X^i$'s, similarly to L\"uscher techniques used to relate
pion scattering phases to lattice data \cite{Luscher:1986pf}.
 Then the three excited energy levels allow to compute the phase shift in three channels.   
  For the minimal Nambu--Goto strings all three states are degenerate at the tree level. One-loop Polchinski--Strominger amplitude splits spin 2 and spin 0 states, but does not lift the degeneracy between the scalar and the pseudoscalar, as shown by dotted lines in Figures ~\ref{fig:plots3} and~\ref{fig:plots5}.  Instead, in the lattice data  the pseudoscalar state is strongly splitted from  other states. The size of this splitting is much larger than the one loop splitting, indicating
 that  a qualitatively new ingredient is required to explain the pseudoscalar state. Given that its energy  is almost independent of $R$ it is natural to interpret this level as arising from a massive pseudoscalar resonance (the worldsheet axion) localized on a string. This interpretation is supported by observing the same boosted particle in higher energy excitations, carrying non-zero total momentum 
along the string.

\begin{figure}[t!]
    \begin{center}
        \includegraphics[width=3.15in]{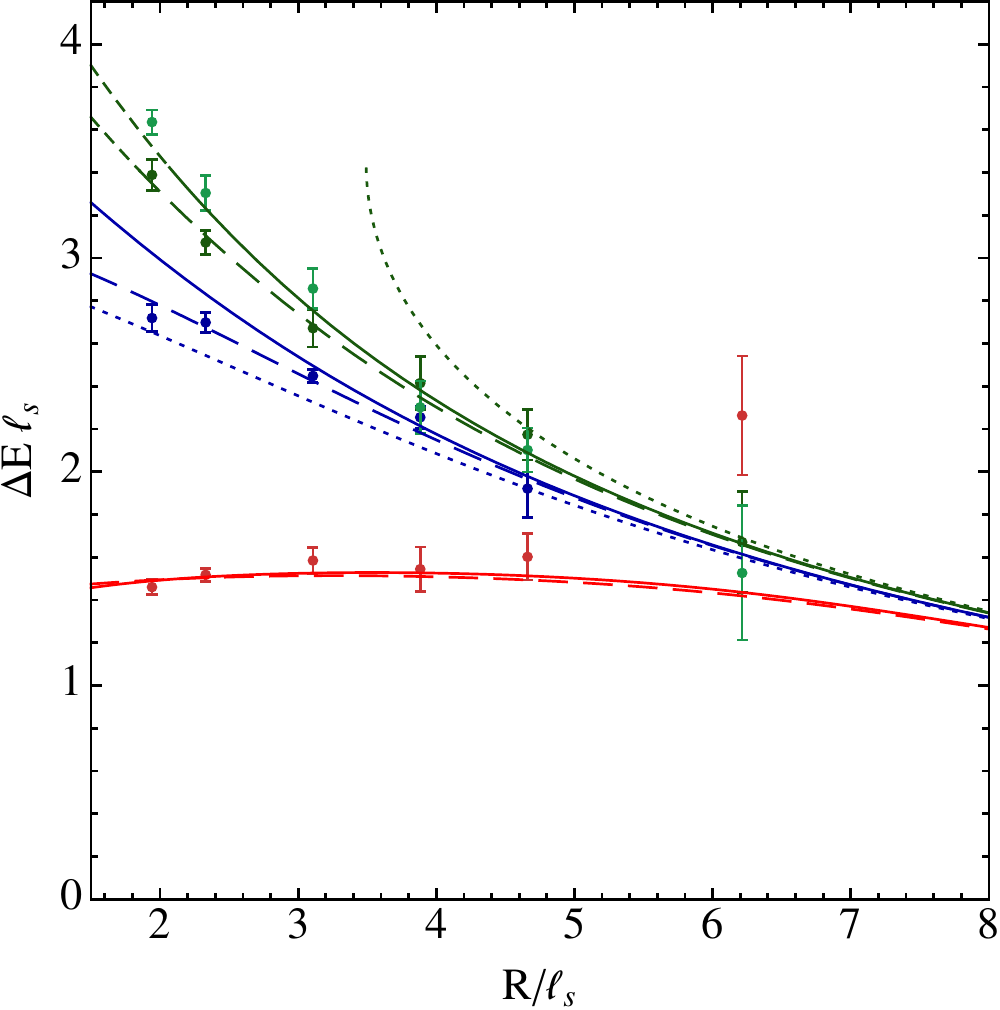}
        \hspace{0.02cm}
          \includegraphics[width=3.15in]{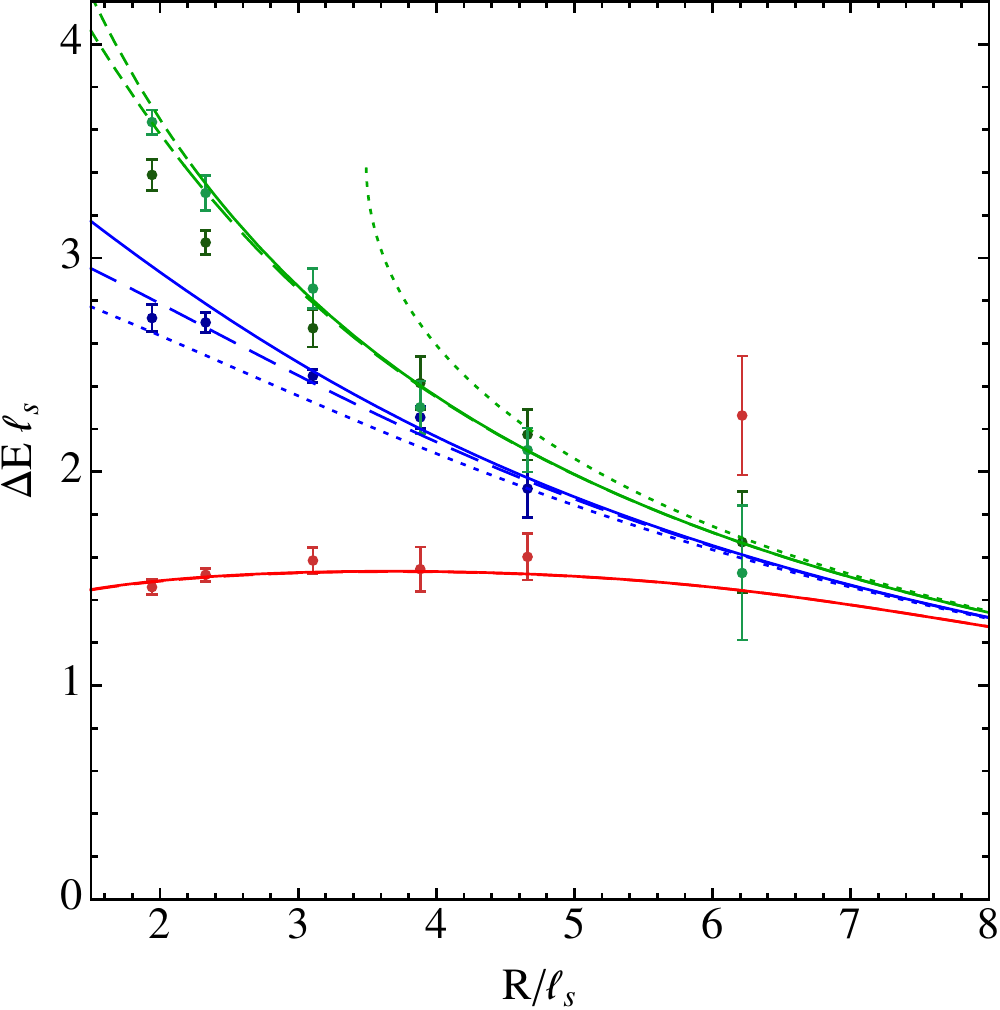}
        \caption{This plot shows $\Delta E=E-R/\ell_s^2$ as a function of  length of the $SU(3)$ flux tube for the two particle states in red, blue, green and dark green for $0^{--}$, $0^{++}$, $2^{+-}$ and $2^{++}$ states respectively. The data is taken from~\cite{Athenodorou:2010cs}. The solid lines show the theoretical predictions derived from phase shift (\ref{delta_res}) with the best fit value corresponding to $2^{++}$ states (left) and $2^{+-}$ states (right). The dashed lines correspond to the fit with higher derivative corrections (\ref{delta2}). Shorter dashing indicates Goldstone momenta above $1.85\ell_s$ where the one loop contribution into the phase shift becomes equal to the tree level one. Dotted lines show theoretical predictions without the resonance.}
         \label{fig:plots3}
    \end{center}
\end{figure}
\begin{figure}[t!]
    \begin{center}
        \includegraphics[width=3.15in]{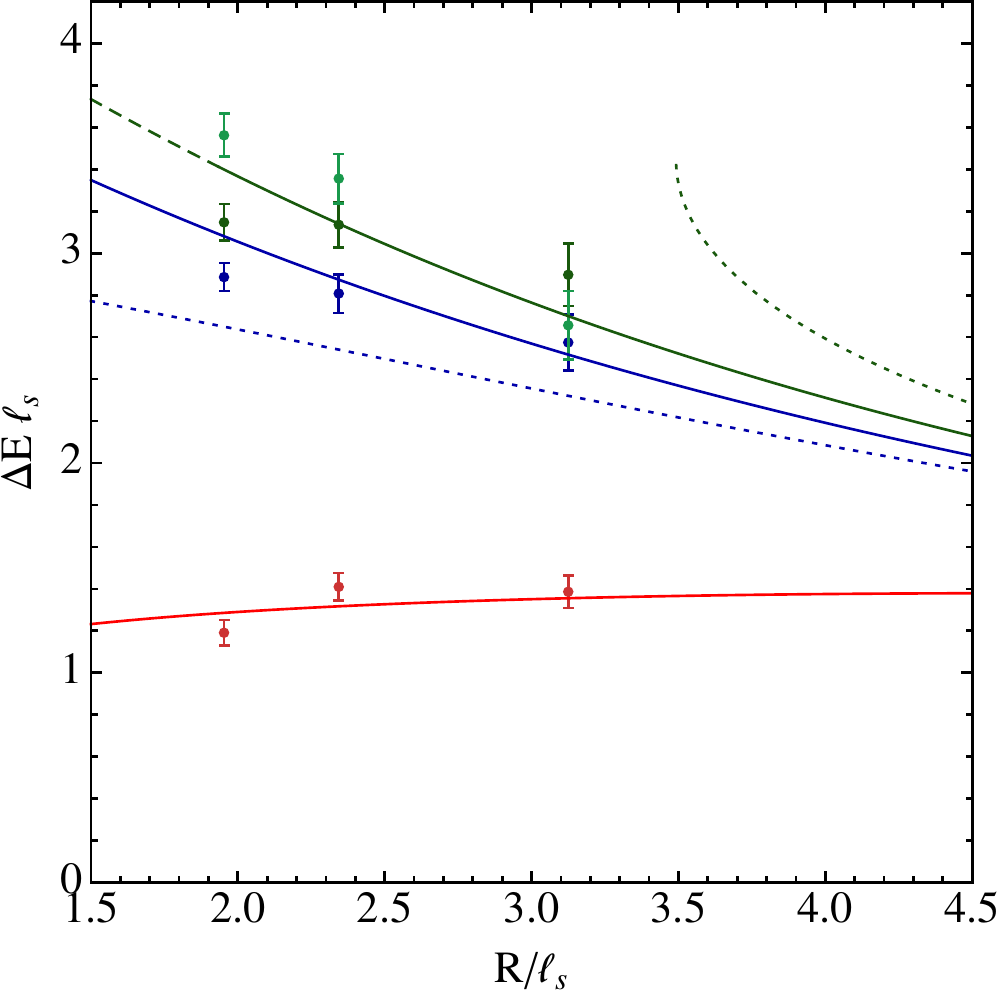}
        \hspace{0.02cm}
          \includegraphics[height=3.15in]{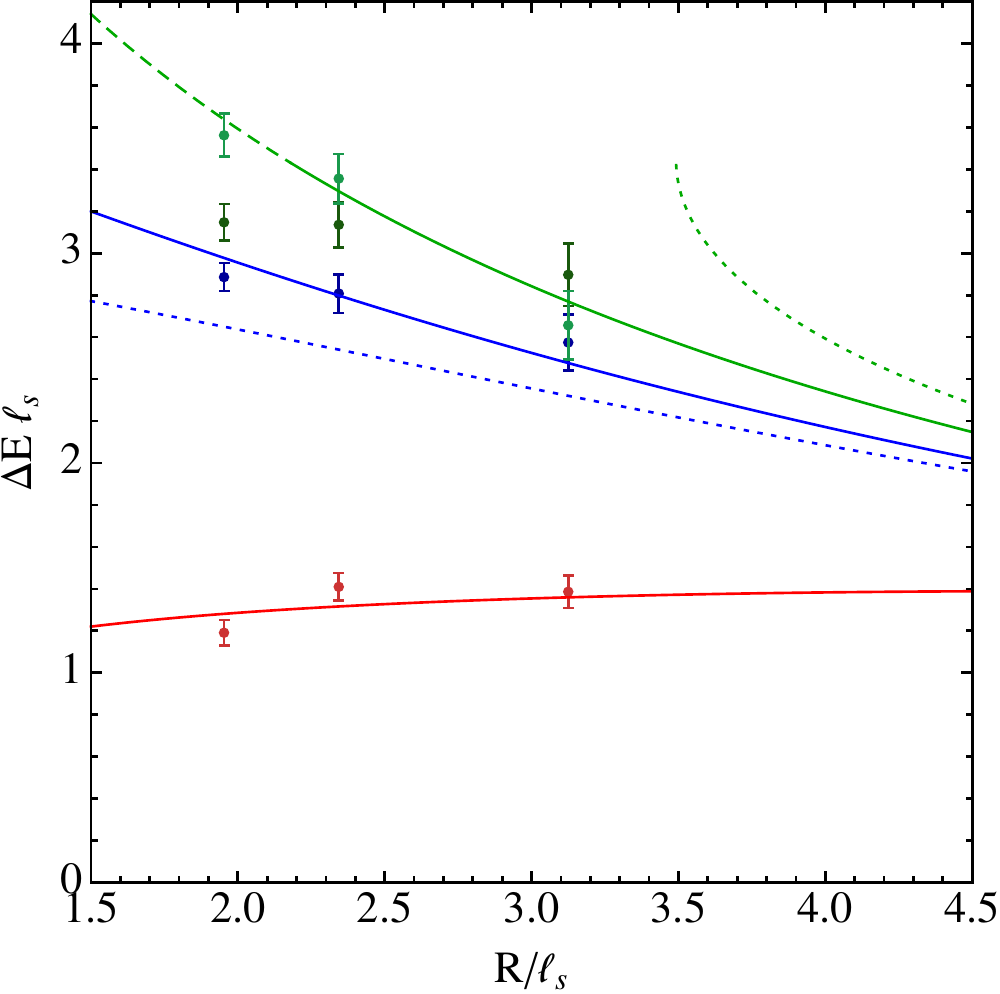}
           \caption{This plot shows $\Delta E=E-R/\ell_s^2$ as a function of  length of the $SU(5)$ flux tube for the two particle states in red, blue, green and dark green for $0^{--}$, $0^{++}$, $2^{+-}$ and $2^{++}$ states respectively. The data is taken from~\cite{Athenodorou:2010cs}. The solid lines show the theoretical predictions derived from phase shift (\ref{delta_res}) with the best fit value corresponding to $2^{++}$ states (left) and $2^{+-}$ states (right). Shorter dashing indicates Goldstone momenta above $1.85\ell_s$ where the one loop contribution into the phase shift becomes equal to the tree level one. Dotted lines show theoretical predictions without the resonance.}
        \label{fig:plots5}
    \end{center}
\end{figure}
\begin{figure}[t!]
    \begin{center}
        \includegraphics[width=3.15in]{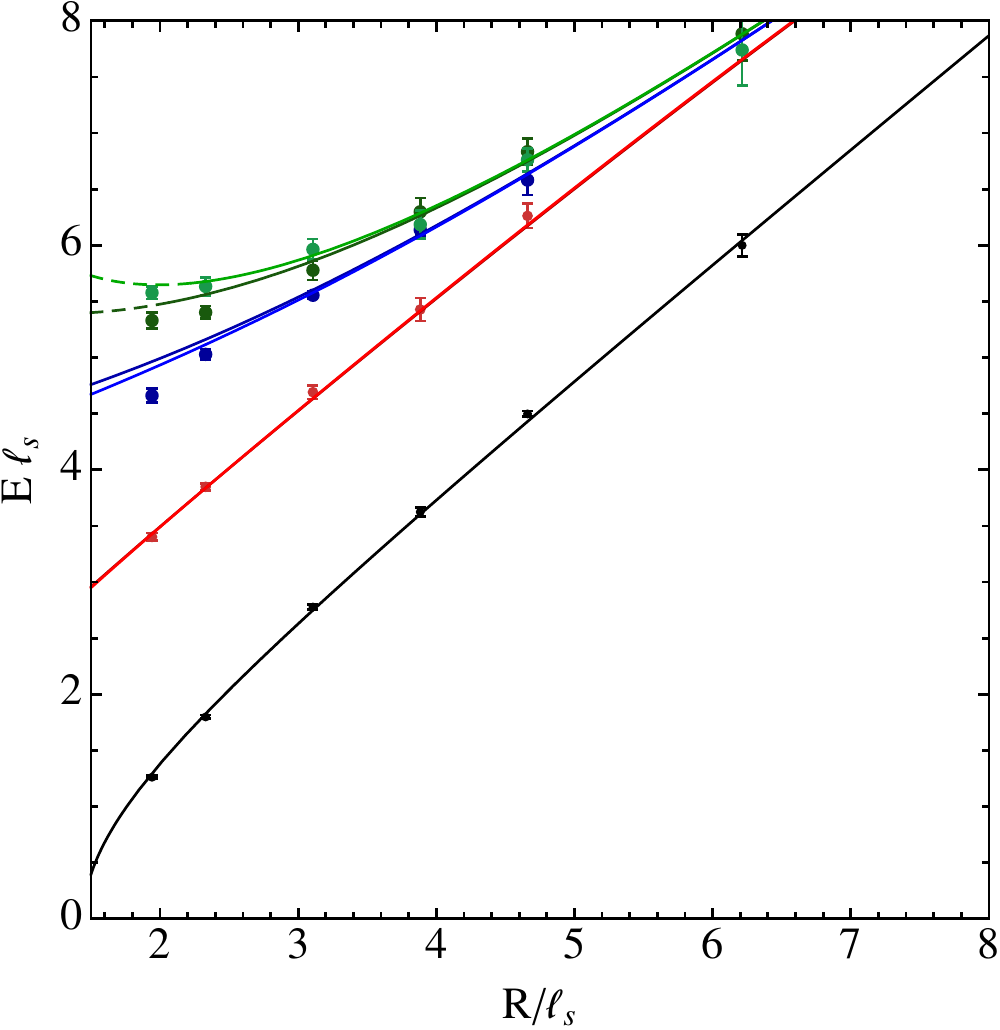}
               \caption{Energy of the $SU(3)$ flux tube as a function of length for the ground state in black and for the two particle states in red, blue, green and dark green for $0^{--}$, $0^{++}$, $2^{+-}$ and $2^{++}$ states respectively. The data is taken from~\cite{Athenodorou:2010cs}. The solid lines show the theoretical predictions derived from phase shift (\ref{delta_res}) with the best fit value corresponding to $2^{++}$ fits (darker colors) and $2^{+-}$ fits (lighter colors). Shorter dashing indicates Goldstone momenta above $1.85\ell_s$, where the one loop contribution becomes equal to the tree level one.}
        \label{fig:plotsC}
    \end{center}
\end{figure}
In the presence of the axion the effective string action is a sum of the Nambu--Goto part and the axion contribution,
\be
S_{massive}=-\int d^2\sigma\sqrt{-h}\l \ell_s^{-2}+(\d a)^2
+\half m^2 a^2\r +S_{a}+\dots,
\ee
where $S_a$ is given by (\ref{S1a}) and $\dots$ stand for terms which contribute to scattering of $X^i$'s only at order $\ell_s^6$ and higher.
The two-particle phase shifts are
\be
\label{delta_res}
2\delta(p)=\ell_s^2 p^2-(2\sigma_2-\sigma_1)\frac{22}{24 \pi}\ell_s^4 p^4+ 2\sigma_2 \tan^{-1}\l{8Q_a^2\ell_s^4{p}^6\over m^2-4{p}^2}\r+ \sigma_1{8Q_a^2\ell_s^4{p}^6\over4{p}^2+m^2}+O(\ell_s^6p^6)\,.
\ee
with $\sigma_1=(-1,1,1)$, $\sigma_2=(0,0,1)$,
for scalar, symmetric, and pseudoscalar channels, respectively. The first term is the tree-level Nambu-Goto phase shift. The second term is the one loop Polchinski--Strominger contribution proportional to $D-26=-22$, the third term represents the resonant $s$-channel contribution, while the last one
arises from the $t$- and $u$-channels.

The next step in the TBA procedure is to plug the phase shift into the TBA equations  and to compare the resulting finite volume spectra with lattice results.
When matching the phase shift to the lattice data, we also obtain momenta of branons comprising the corresponding state. At low enough momenta
  two loop terms can be neglected and  the phase shift in all three channels depends only on two parameters: $m$ and $Q_a$. For data points corresponding to shorter strings higher-derivative terms start playing a role. Since there are local Lorentz-invariant terms in the action of order $\ell_s^6$ at this order the phase shift is not universal and those terms result into the major theoretical uncertainty in the lattice measurement of the coupling $Q_a$.

 The mass of the resonance  $m$ is essentially determined by the pseudoscalar state alone. On the other hand,  changing $Q_a$  mostly affects scalar and tensor energies due to cross-channel terms in (\ref{delta_res}). In particular, to determine $Q_a$ we cannot ignore the spurious splitting between the two tensor states $2^{++}$ and $2^{+-}$. In order to address this issue we do two separate fits,
 one using  $2^{++}$ data only, and the second one with $2^{+-}$ alone{\footnote{In \cite{Dubovsky:2013gi,Dubovsky:2014fma} only the $2^{++}$ state was used for the fit.}}.
\begin{table}[t!]
\begin{center}
 \begin{tabular}{ | c  | c | c | }
   \hline			
     & $SU(3)$ & $SU(5)$ \\
    \hline
  $2^{++}$ &  &   \\
  $m\ell_s$ & $1.85^{+0.02}_{-0.03}$ & $1.64^{+0.04}_{-0.04}$ \\[6pt]
   $Q_a$ & $0.380^{+0.006}_{-0.006}$ & $0.389^{+0.008}_{-0.008}$ \\ [6pt]
   \hline  
   $2^{+-}$ &  &   \\
  $m\ell_s$ & $1.85^{+0.02}_{-0.02}$ & $1.64^{+0.04}_{-0.04}$ \\[6pt]
   $Q_a$ & $0.358^{+0.004}_{-0.005}$ & $0.358^{+0.009}_{-0.009}$ \\ [6pt]
   \hline  
 \end{tabular}
 \hspace{0.7in}
 \label{table:tab2}
 \begin{tabular}{ | c | c | }
    \hline
  $2^{++}$ &     \\
  $Q_a$ & $0.425$  \\ [4pt]
  $A$ & $1.42$ \\ [4pt]
    $B$ & $1.70$ \\ [4pt]
   \hline  
   $2^{+-}$ &     \\
  $Q_a$ & $0.357$  \\ [4pt]
   $A$ & $0.59$  \\ [4pt]
    $B$ & $0.47$  \\ [4pt]
   \hline  
 \end{tabular}
  \caption{The left table shows the best fit values and statistical uncertainties for parameters $m$ and $Q_a$ in  (\ref{delta_res}). The right table shows the best fit $SU(3)$ values when the higher derivative corrections (\ref{delta2}) are included.}
      \label{tab:tab1}
  \end{center}
\end{table}

The results are presented in Table 1.
As expected, the value of $m$ is independent of the choice of a tensor state.
The value of  $Q_a$ varies somewhat but remains within $4$\% around (\ref{QI}). This split may be taken as an estimate of a systematic uncertainty.

To estimate theoretical uncertainties we add a contribution to the phase shift that could be caused by  local two-loop terms in the action. There are two counterterms at ${\cal O}(\ell_s^6)$.
By denoting the corresponding coefficients  $A$ and $B$ the correction to the phase shift can be parametrized as
\be
\label{delta2}
2\delta_{2-loop}(p)={1\over 4\pi^2}\l \sigma_3 A -B \r \ell_s^6 p^6,
\ee
where, as before, $\sigma_3=(3,1,-1)$ for scalar, symmetric, and pseudoscalar channels. 
We included the two loop phase space factor $(2\pi)^{-2}$, so that one expects $A$ and $B$ to be order one. 
Note that we did not include higher derivative corrections to the cubic vertex (\ref{S1a}) of the form
\be
\ell_s^4\int d^2\sigma a \epsilon_{ij}\epsilon^{\alpha\beta}\d_\alpha\d_\gamma \d_\sigma X^i\d_\beta\d^\gamma \d^\sigma X^j+\dots\;
\ee
because by using integrations by parts and field redefinitions they can be reduced to the change in the couplings $Q_a$, $A$ and $B$.
 A fit to the $SU(3)$ data with these two new additional parameters results in
central values for $Q_a$, $A$ and $B$ presented in  Table \ref{tab:tab1}. 
We see that the best fit value of $Q_a$ is practically unaffected for the fit involving the $2^{+-}$ tensor state, and shifts 
by $\sim 12$\% for the fit involving the $2^{++}$ state.  
We take this shift as an estimate of the theoretical uncertainty.  Due to the lack of data points we did not perform a fit involving two-loop terms for $SU(5)$ data.

Note that this shift is to large extent driven by a single data point corresponding to the energy of the shortest string in the scalar channel, which is 
not fitted well by the two parametric model with $m$ and  $Q_a$ alone. Several factors may contribute to this, such as possible large winding corrections associated with a new pseudoscalar state and not included in the TBA equations used in  \cite{Dubovsky:2013gi,Dubovsky:2014fma}.
Of course, it is also very reasonable to expect higher order corrections to the phase shift at the level corresponding to the best fit values of $A$ and $B$. The most straightforward way to reduce theoretical uncertainties is to impose a cut on the length of the strings used in the fit.
We checked that excluding the shortest strings from the fit indeed allows to  significantly suppress the theoretical uncertainty. Of course, this also results in larger statistical error bars, which still remain below $10$\%. Overall, we feel it is safe to conclude from this analysis that $Q_a$ and $Q$  agree 
at the $\sim10$\% level.

\section{Discussion and future directions}
\label{sec:last}
To summarize, we presented a construction of a new  family of non-critical relativistic strings with integrable dynamics on the worldsheet. In particular, this family includes a four-dimensional string theory with a single new massless  axion mode on the worldsheet. Surprisingly, the coupling constant of a massive axion present on the worldsheet of confining strings in pure gluodynamics agrees  with a 10\% precision with the value required for integrability.

These results raise two general classes of questions. The first one, which is not directly related to the dynamics of gauge theories,  is what members of this family can be promoted to interacting string theories, and what is the interpretation of this family (if any) from the critical string theory viewpoint. These questions are partially understood for linear dilaton non-critical strings, which is the subset of our family, which have a known  Polyakov 
formulation with a tractable CFT. Given the important role played by integrability in much of the recent progress in string theory, we feel answering these questions should be rewarding.

The second set of questions is more directly related to gluodynamics. These include understanding of what is the proper interpretation of the numerical coincidence which we observed. Another important question in this category is what are the next steps to be done both on theoretical and ``experimental" (lattice) side to study the implications of $Q_a\approx Q$.

The most optimistic interpretation of our results would be that they hint that the confining string in planar gluodynamics is described by an integrable axionic string.
For this to be possible, the axion mass should vanish in the infinite $N_c$ limit. Comparing the $SU(3)$ and $SU(5)$ data presented in Table \ref{tab:tab1}, one finds that the axion mass does go down a bit. The shift is not too large. However, it should not be ignored, especially given that the value of $Q_a$ remains the same within the error bars. In any case, a clear advantage of this scenario is that it has a definite prediction which should be possible to check with the future lattice data.
 
 A more conservative, but still very interesting possibility, is that the axion mass does not vanish in the large $N_c$ limit, but the high energy dynamics of the QCD confining string is governed by the integrable axionic theory.
 
 Finally, the most pessimistic option is that the coincidence observed here is purely numerological and does not teach us anything about confining strings. Indeed, a sceptic may wonder what would be a reason for the QCD string to give rise to a particularly nice and interesting structures, such as integrability. There are many four-dimensional theories exhibiting string-like topological defects, such as the Abelian Higgs model, and in general there is no reason to find an integrable theory on the world-sheet of these strings. 
 
 We believe this concern is to large extent addressed by the arguments presented in the {\it Introduction}. What distinguishes the QCD strings from  many other topological defects is the existence of the planar limit, in which one expects to obtain a UV complete theory on the string worldsheet. On geometrical grounds, this theory is expected to exhibit time delays growing linearly with the collision energy and to exhibit asymptotic fragility. Given  that this theory is coming from an asymptotically free theory in the bulk, one may expect it to be somewhat simple, and the minimal option characterized by the $S$-matrix (\ref{S-matrix}) does not sound unreasonable (at least, as the UV limit).
Even if this is not the case, we see that the third option is not pessimistic at all, as we are still left with an interesting and challenging question to characterize the dynamics of this two-dimensional theory.

The good news are that the combination of lattice simulations with the TBA technique provides us with a sufficient toolkit to make progress, and in particular to distinguish between the second and the third option.
A clear improvement in this direction would be achieved by increasing the  quality of lattice data for the levels studied in section~\ref{sec:data} for intermideate lengths $R/\ell_s\sim 2.5\div 5$, 
by extending the analysis to a larger number of colors, and by getting rid of the systematic uncertainty caused by the spurious $2^{++}$ and $2^{+-}$ splitting. The latter problem should be possible to solve by changing the shape of a lattice (e.g., by using the hexagonal lattice). As mentioned at the end of section~\ref{sec:data}, this will allow to decrease the theoretical uncertainty in measuring the value of $Q$ by imposing the stronger cut at the short 
string end.

An even stronger progress will be achieved when the high quality data for higher excited string states becomes available in the same range of $R/\ell_s$.  Likely, the most reasonable strategy in this case will be to use the TBA equations for measuring the two-dimensional $S$-matrix, rather than to fit the finite energy spectra as we are doing in the present paper. The prediction following from the second option is that the resulting $S$-matrix should approach (\ref{S-matrix}) at high energies and large $N_c$.  Of course, these data will only be useful for excitation energies below the cutoff scale $\Lambda_{N_c}$, where two-dimensional unitarity breaks down.
One can estimate the scaling of $\Lambda_{N_c}$ with $N_c$ in the following way. Let us consider a colliding pair of generic left- and right-moving excitations each of energy $E$ on the flux tube worldsheet. Semiclassically, the   production of the bulk states proceeds as a result of string self-intersections in the course of collisions, with subsequent string interconnection. Then the number of self-intersections per unit time per unit length scales as  \cite{Polchinski:2006ee}
\[
d{\cal N}\propto {dl\over l^3}\;,
\]
where $l$ is a length of a loop produced as a result of a self-intersection. 
This result implies that most of self-intersections happen at short scales and their frequency is $\dot{{\cal N}}\sim\ell_s^{-2}$.
Then the total number of self-intersections is of order
\[
{\cal N}\sim \dot{ {\cal N}} L^2\sim E^2\ell_s^2\;,
\]
where $L\sim \ell_s^2 E$ is both the characteristic size of the collision region and the characteristic duration of the collision.
The probability of interconnection in the course of a self-intersection is proportional to $g_s^2\sim N_c^{-2}$, where $g_s$ is the string coupling constant.
Hence, the probability of a glueball production is given by
\[
P_{gl}\sim {E^2\ell_s^2\over N_c^2}\;.
\]
It becomes large at the scale
\[
\Lambda_{N_c}\sim {N_c\over \ell_s}\;.
\]
Note that for fundamental strings this would be the relation between the string and the Planck scales.

To conclude,  understanding the dynamics of the planar QCD string remains an outstanding and fascinating question. We think that the results presented in the present paper support an idea that this question may also have a fascinating answer, and this answer may perhaps be simpler than what one might expect {\it a priori}. At any rate, the TBA technique combined with lattice simulations opens a direct path towards making a concrete progress in resolving this question. We are optimistic that the progress will be achieved in the near future.

\section*{Acknowledgements}
We would like to thank Raphael Flauger for collaboration on many related topics.
We are also grateful to  Nima Arkani-Hamed, Andreas Athenodorou, Simeon Hellerman, Guzm\'an Hern\'andez-Chifflet, Vladimir Kazakov, Markus Luty, Mehrdad Mirbabayi, Mike Teper, V.P. Nair, Eva Silverstein, Cobi Sonnenschein and many others for stimulating discussions.
This work was supported in part by the NSF CAREER award PHY-1352119.
\section{Appendix: From double softness to the Galilean symmetry}
\label{sec:app}
The purpose of this Appendix is to justify  the statement, which we relied on in section~\ref{sec:uniq}.  The statement is the following.
A local two-dimensional vertex $V(X^i)$ with more than four legs, which produces double soft on-shell amplitude w.r.t. to any incoming momentum, can be 
written in the Galilean invariant
form ({\it i.e.}, invariant under (\ref{Galilean}) for any $i$). Here we allow to make use of the field equations, {\it i.e.} to add/subtract terms which vanish on-shell.

The argument consists of two steps. First, let us prove a weaker statement, namely that double softness of the on-shell amplitude implies Galilean symmetry of $V$ when all fields $X^i$ are on-shell. The fastest 
way to check this is to go through the steps required to calculate the on-shell amplitude corresponding to $V$.
Namely, one writes the field as the sum of the Fourier modes
\[
X^i=\sum a^i_\alpha e^{i p^i_\alpha\sigma}\;,
\]
where $p^i_\alpha$ are physical on-shell momenta. Then one plugs this expression into the vertex and evaluates the result at linear order in all $a^i_\alpha$. The corresponding coefficient is the on-shell amplitude. Double softness in one of the momenta 
$p^i_\alpha$ implies that if one treats $p^i_\alpha$ as a Grassmann variable, the corresponding amplitudes will all be zero.
This is equivalent to the on-shell Galilean invariance of $V$, because for a Grassmann momentum one may replace,
\[
e^{i p^i_\alpha\sigma}=1+ip^i_\alpha\sigma\;.
\]
Note that restricting to a linear order in $a^i_\alpha$'s does not invalidate the argument, because by probing a sufficiently large   set of on-shell configurations, invariance at linear order in all $a^i_\alpha$'s implies  invariance of the full vertex $V$.

An alternative proof of this statement uses the fact that the shift of some field under the Gallilean symmetry is equivalent to convolution of the vertex with a wave function equal to $\delta'(p)$. Now, if in the momentum space the vertex reads $V(p,\{q_i\})\delta(p+\sum q_i)$, then the variation under the Gallilean shift is proportional to
\be
\delta'(p)V(p,\{q_i\})\delta\l p+\sum q_i\r\sim V(0,\{q_i\})\delta'\l\sum q_i\r+\frac{\d}{\d p} V(0,\{q_i\})\delta\l\sum q_i\r=0\;,
\ee
where the first and the second terms vanish due to softness and double-softness of $V$ correspondingly.

Let us now demonstrate that {\it on-shell} Galilean invariance of $V$ implies that it can be rewritten in the {\it off-shell} invariant form by adding and subtracting terms vanishing on-shell, provided $V$ has more than four legs. 
This part of the argument will be specific to two dimensions. Namely, after dropping terms which vanish on-shell, a local two-dimensional  vertex $V$ can be presented in the following factorized form,
\[
V=\int dx^+dx^-\sum a_{AB}V_-^A(\d_-X)V_+^B(\d_+X)\;, 
\]
where to make the argument shorter we made use of the shift invariance of the vertex $V$, which is one of the starting assumptions in the argument of section~\ref{sec:uniq}.
In principle, it is straightforward to relax this assumption, because it also follows from the double softness. Note, that in general shift invariance does not imply that every field is acted upon by a derivative, one may also have Wess--Zumino terms of the form $V^{WZ}_-=X\d_-Y$ for some field $X$, $Y$. However, these are forbidden by double softness. 

Now, on-shell Galilean invariance of $V$ implies off-shell Galilean invariance of one-dimensional ``vertices" $\int dx^-V_-^A(\d_-X)$, $\int dx^+V_+^B(\d_+X)$.  This is satisfied either if $V_\pm$ are trivial invariants, {\it i.e.}, contain at least two derivatives acting on every field, $V_\pm=V_\pm(\d_\pm^2X,\dots)$, or if they are Wess--Zumino terms of one of the following two structures,
\begin{gather}
\nonumber
V_+^{WZ1}=\d_+ X\d_+Y\\
\nonumber
V_+^{WZ2}=\d_+ X\d^2_+Y\;,
\end{gather}
and the same with $+\leftrightarrow -$. Consequently, apart from the two double soft quartic vertices with four and six derivatives $V_+^{WZ1}V_-^{WZ1}$ and $V_+^{WZ2}V_-^{WZ2}$, all other vertices are Galilean invariant off-shell and, moreover, have at least two derivatives acting on each field, which is exactly the statement we need in section~\ref{sec:uniq}.

It appears very plausible that some version of the statement which we just proved holds also for higher order softness and in higher dimensions. Note also, that most likely the need for a somewhat lengthy and clumsy argument for such an intuitive statement, is not entirely due to our inability to find a more elegant proof. For instance, already in two dimensions in the absence of the (worldsheet) Lorentz symmetry we would be able to construct a much larger set of double soft vertices, which are not invariant under the Galilean symmetry. For instance, any vertex of the form 
\[
\int dx^+dx^- V_+^{WZ1,2}V_-(\d_-^2X)
\] 
would do the job.


\bibliographystyle{utphys}
\bibliography{dlrrefs}
\end{document}